\documentstyle[12pt]{article}
\newcommand{\beq}{\begin{equation}}
\newcommand{\bea}{\begin{eqnarray}}
\newcommand{\eeq}{\end{equation}}
\newcommand{\eea}{\end{eqnarray}}
\newcommand{\bcen}{\begin{center}}
\newcommand{\ecen}{\end{center}}
\newcommand{\bM}{\overline{M}}

\newcommand{\fp}{{\mathrm{FP}}}
\newcommand{\ut}{u_t^{\mathrm{FP}}}
\newcommand{\ub}{u_b^{\mathrm{FP}}}
\newcommand{\utau}{u_{\tau}^{\mathrm{FP}}}
\newcommand{\ul}{u_{\lambda}^{\mathrm{FP}}}

\newcommand{\et}{e_t^{\mathrm{FP}}}
\newcommand{\eb}{e_b^{\mathrm{FP}}}
\newcommand{\etau}{e_{\tau}^{\mathrm{FP}}}
\newcommand{\el}{e_{\lambda}^{\mathrm{FP}}}
\newcommand{\ek}{e_{\kappa}^{\mathrm{FP}}}
\newcommand{\xt}{\xi_t^{\mathrm{FP}}}
\newcommand{\xb}{\xi_b^{\mathrm{FP}}}
\newcommand{\xtau}{\xi_{\tau}^{\mathrm{FP}}}
\newcommand{\xl}{\xi_{\lambda}^{\mathrm{FP}}}
\newcommand{\xk}{\xi_{\kappa}^{\mathrm{FP}}}
\newcommand{\lsim}{\raisebox{-0.13cm}{~\shortstack{$<$ \\[-0.07cm] $\sim$}}~}
\newcommand{\gsim}{\raisebox{-0.13cm}{~\shortstack{$>$ \\[-0.07cm] $\sim$}}~}
\newcommand{\eq}[1]{\mbox{(\ref{eq:#1})}}
\newcommand{\defprod}{\raisebox{-0.13cm}{~\shortstack{$\prod$ \\[-0.07cm] 
${}_{j\neq k}$}}~}

\def\tvi{\vrule height 12pt depth 6pt width 0pt}
\def\tv{\tvi\vrule}
\def\cc#1{\kern .7em\hfill #1 \hfill\kern .7em}

\begin{document}
\begin{flushright}
PM/00--46 \\
%GDR--S--??\\
hep-ph/0101237
\end{flushright}
\begin{center}
{\Large \bf General one-loop renormalization group evolutions

 and electroweak symmetry breaking\\[.25cm]
 in the (M+1)SSM }

\vspace{1cm}

{\sc  Y. Mambrini\footnote{{\sl Allocataire} MNESR}, 
G. Moultaka, M. Rausch de Traubenberg\footnote{On leave of absence from {\sl Universit\'e  Louis Pasteur, Strasbourg}.} }

%\vspace{0.5cm}
{\it$^b$ Physique Math\'ematique et Th\'eorique, UMR No 5825--CNRS, \\
Universit\'e Montpellier II, F--34095 Montpellier Cedex 5, France.
}
\end{center}
\vspace{1cm}

\begin{abstract}
%In this paper, we give the analytical form of the solution of the RGE in the
%frame of the (M+1)SSM, for the Yukawa and the soft terms. We also study
%different Fixed Point Regimes, and exclude some phenomenologically.

We study analytically the general features of electroweak symmetry 
breaking in the context of the Minimal Supersymmetric Standard 
Model extended by one Higgs singlet. 
The exact analytical forms of the renormalization group evolutions of 
the Yukawa couplings and of the soft supersymmetry breaking parameters are 
derived to one-loop order. They allow on one hand controllable approximations
in closed analytical form, and on the other a precise study of the
behaviour of infrared quasi fixed point regimes which we carry out. 
Some of these regimes are shown to be phenomenologically inconsistent, leading 
to too small an effective  $\mu$-parameter. 
The remaining ones serve as a suitable benchmark  to understand 
analytically some salient aspects, often noticed numerically in the literature,
 in relation to the electroweak symmetry breaking in this model. 
The study does not need any specific assumption on $\tan \beta$ or 
on boundary conditions for the soft supersymmetry breaking parameters, 
thus allowing a general insight into the sensitivity of the low energy physics 
to high energy assumptions.          
 
\end{abstract}
\newpage
\section{Introduction.}

Extensions of the Higgs sector of the standard model or of the minimal supersymmetric standard model (MSSM) \cite{MSSM} , 
is a suitable framework to assess the
 phenomenology of the search for Higgs-like particles, and in a wider context, 
that of the supersymmetric partners of these particles. The next to MSSM,
dubbed hereafter (M+1)SSM \cite{Nilles, Derendinger, Ellis, EH0, Binet},
 where one $SU(3)_c \times SU(2)_L \times U(1)_Y$ gauge
singlet supermultiplet is added \cite{Fayet}
 to the MSSM, has attracted interest, 
initially as a framework for a natural solution to the so-called $\mu$-problem
\cite{Nilles}  
and later on as a source for interesting phenomenology which can differ from 
that of the MSSM. 
Due to the modification of the Higgs sector, the phenomenology and the upper 
bound on the mass of the lightest (observable) Higgs is altered \cite{EH1},
while the modified neutralino sector can lead to unconventional signatures 
for the sparticle searches \cite{EH2}. 

Our main concern in this paper is the pattern of electroweak symmetry breaking
(EWSB)
and the dynamical generation of the $\mu$ parameter.
As was initially noted in \cite{ERS1} and  studied intensively 
\cite{ERS1, ERS2}, the v.e.v. of the gauge singlet field tends to be much 
larger than those of the two Higgs doublets, implying the tendency for the 
singlet chiral superfield to  decouple from the
 other superfields of the theory. This means that, apart from 
some ranges of the parameter space where the singlino is the lightest 
supersymmetric particle, the features of the MSSM are basically shared by its 
minimal extension.
However, the analysis of the issues in (M+1)SSM was mostly done numerically,
or, when analytically, only for small $\tan \beta$ \cite{ERS2}
(no Yukawa couplings for the tau and bottom were considered). Furthermore this
is often done in the framework of universality of the soft supersymmetry (susy) 
breaking parameters, or in some scenarios with mild non-universality
which gave results comparable to the ones with universality \cite{Brax}.

In view of the importance of such features (and possibly their generalization
to an extended singlet sector), it is important to attempt an
understanding of the generic pattern for the dynamical determination
of the v.e.v. of the singlet, without any specific assumptions about
the GUT-scale boundary conditions or the magnitude of $\tan \beta$.
In this paper we will address this issue fully analytically and from
two complementary sides: firstly the analytical evolution of the various
Yukawa (and gauge) couplings and the soft susy breaking parameters,
as well as the determination of the regimes of least sensitivity
to initial conditions, {\sl i.e.} in the vicinity of 
the infrared quasi fixed points\footnote{
another possibility of least sensitivity, not considered here,
could be the occurrence of focus points at phenomenologically acceptable energy 
scales much like in the MSSM\cite{Feng}. }(IRQFP); 
secondly the study of the  EWSB equations in compelling regimes and the 
interplay between the magnitude of the dynamically determined Higgs doublet
mixing parameter $\hat{\mu}$, the experimental lower bounds 
on chargino masses, and the relaxation of universality of the soft parameters.

The paper is organized as follows: 
In section 2 we recall the basic ingredients of the (M+1)SSM
and introduce our notations. In section 3 we deal with the  renormalization 
group equations (RGEs), give the analytical integrated forms
of their solutions along lines similar to \cite{Auberson, Kazakov} and classify the ensuing 
IRQFPs regimes. Four regimes are found generalizing the MSSM case 
\cite{Moultaka}. 
Some numerical illustrations of the IRQFPs regimes are given in
section 4. Section 5 is devoted to the study of EWSB constraints and
we conclude in section 6. More technical derivations and results are
given in Appendices.  

 From a different standpoint, it is worth keeping in mind potential difficulties
that can arise in the (M+1)SSM in relation to the appearance of cosmologically 
problematic domain-wall solutions. We, however, bypass in this paper such 
problems and possible solutions to them \cite{Abel}.  
Another interesting feature of the (M+1)SSM on which we will not dwell is the 
possibility to break spontaneously CP symmetry. In this paper we will
will assume, without further reference, to be in regions of parameter space 
where CP is broken neither explicitly nor spontaneously \cite{Romao}.

\renewcommand{\theequation}{1.\arabic{equation}}
\setcounter{equation}{0}
\section{The (M+1)SSM.}
\renewcommand{\theequation}{2.\arabic{equation}}
\setcounter{equation}{0}

In this model,
the Higgs sector is constituted by two Higgs doublets $H_1$ and $H_2$, 
and one singlet $S$.
The spectrum, compared to MSSM is richer (one more CP even, CP odd
Higgs field, and one more neutralino). Introducing as in the MSSM
the matter fields ($Q$, $T$, $B$, $E$, $L$) the superpotential 
reads 

\beq
W={\lambda}\hat S \hat H_1. \hat H_2+
\frac{{\kappa}}{3} \hat S^3 +y_t \hat T \hat Q. \hat H_2 
+y_b \hat B \hat Q. \hat H_1
+y_{\tau} \hat E \hat L. \hat H_1
\label{eq:supot}
\eeq

\noindent
where the dot product represents the SU(2) scalar product,
and the superfields $\hat T$ ($\hat B$) respectively, the left 
handed antitop(antibottom), $\hat E$, the left-handed antitau, 
and $\hat Q$ ($\hat L$) the left handed doublets for quarks(lep\-tons).
All the parameters in the superpotential are  dimensionless,
and a  mass term is forbidden by a discrete $Z_3$ symmetry.
This symmetry also prevents $S$ to take large v.e.v. ($<S>$) and
an effective $\mu$ parameter is 
generated ($\hat{\mu}={\lambda}<S>$) of the order of magnitude
of $100$ $GeV$. But, when $S$ develops a v.e.v.
 this discrete symmetry is broken, and 
domain wall solutions appear. It is well known that such topological 
defects are excluded by cosmology. Possible solutions to this
are given in \cite{Abel}.\\

Supersymmetry breaking is parameterized by the so-called soft SUSY 
terms involving trilinear couplings ($A$'s), scalar and gaugino masses
($m$'s and $M$'s). In \eq{supot} $\hat H_1, \hat H_2$ etc.  
represent the superfields and now, although $H_1, H_1$ etc 
 represent its scalar
component, $\lambda_1$, $\lambda_2$ and $\lambda_3$ are the
$U(1)_Y$, $SU(2)_L$ and $SU(3)_c$ gauginos respectively.

\bea
{\cal L}_{{\mathrm soft}}&=&
M_1\lambda_1\lambda_1+M_2\lambda_2\lambda_2+M_3\lambda_3\lambda_3
+(\lambda A_{\lambda}SH_1.H_2
+\frac{\kappa}{3}A_{\kappa}S^3
\nonumber \\
&&
+ y_t A_t T Q.H_2 +y_b A_b B Q.H_1
+y_{\tau} A_{\tau} E L.H_1
+ \mathrm{h.c.} 
)
\nonumber
\\
&&
+m_1^2|H_1|^2
+m_2^2|H_2|^2
+m_S^2|S|^2
+m_Q^2|Q|^2 \nonumber \\
&&
+m_T^2|T|^2
+m_B^2|B|^2
+m_E^2|E|^2
+m_L^2|L|^2
\label{eq:lsoft} 
\eea

Finally, let us write for later reference, the tree-level scalar potential in 
the neutral Higgs sector including the F-terms from Eq.\eq{supot}, the D-terms 
and the contributions from the soft terms Eq.\eq{lsoft},

\begin{eqnarray}
V&=& m_1^2 |H_1^0|^2 + m_2^2 |H_2^0|^2 + m_s^2 |S|^2 + \frac{\bar{g}^2}{4} ( |H_1^0|^2 - |H_2^0|^2)^2 \nonumber \\
&& + |\kappa S^2 + \lambda H_1^0 H_2^0|^2 + \lambda^2 |S|^2 (|H_1^0|^2 + |H_2^0|^2)  \nonumber \\
&& + (A_{\lambda} \lambda S H_1^0 H_2^0 + A_{\kappa} \frac{\kappa}{3} S^3 + \mathrm{h.c.}) 
\label{eq:scalPot}
\end{eqnarray} 

where $ \bar{g} \equiv \sqrt{(g_1^2 + g_2^2)/2}$, $g_1, g_2$ being the gauge 
couplings associated respectively to $U(1)_Y, SU(2)_L$
and we consider only real valued couplings and mass parameters.

\section{Analytical solution for the renormalization group equation.}
\renewcommand{\theequation}{3.\arabic{equation}}
\setcounter{equation}{0}

\subsection{Exact evolutions to one-loop order}

Using the notations 
$\alpha_i=\frac{g_i^2}{16 \pi^2}$,
 $i=1,2,3$; $Y_j=\frac{y_j^2}{16 \pi^2}$,
 $j=t, b$, $\tau$,
and $Y_{(\lambda,\kappa)}=\frac{(\lambda^2,\kappa^2)}{16 \pi^2}$,
where $g_1,g_2$ and $g_3$ denote respectively the $U(1)_Y,SU(2)_L$ and
 $SU(3)_c$ gauge coupling constants,
one can write down the one-loop RG equations as \cite{Derendinger, Falck}

\begin{eqnarray}
\label{eq:rge}
\dot{\alpha_i}&=&-b_i\alpha_i^2, \nonumber \\
\dot{M_i}&=&- b_i\alpha_i M_i,  \nonumber \\
\dot{Y_k}&=&Y_k(\sum_ic_{ki}\alpha_i-\sum_la_{kl}Y_l) \label{yukeq}\\
\dot{A_k}&=&-\Big(\sum \limits_{i}c_{ki}\alpha_iM_i+\sum \limits_{i}a_{ki}
Y_iA_i \Big)  \nonumber \\
\dot{\Sigma_k}&=&2 \sum\limits_{i}c_{ki}\alpha_iM_i\bM_i
-\sum\limits_{i}a_{ki}Y_i(\Sigma_i+A_i\overline{A_i}) \nonumber 
\end{eqnarray}

\noindent
where $k=t, b, \tau, \lambda, \kappa$,  . $\equiv$ d/dt, t=log $M^2_{GUT}/Q^2$, 
the numerical coefficients $a$'s, $b$'s and
$c$'s are given in Appendix A, and the $\Sigma_k$ are defined in
 Eq.\eq{defSigma}.

Though the RGEs for the Yukawa couplings Eqs.(\ref{yukeq}) do not have explicit analytic 
solutions, they can be solved iteratively as it has been demonstrated 
in \cite{Auberson} through the use of some auxiliary functions.

Together with the gauge couplings, the general solutions for the Yukawa 
couplings read \cite{Auberson}
\begin{eqnarray}
\alpha_i=\frac{\alpha_i^0}{1+b_i\alpha_i^0t} \nonumber \\
Y_k=\frac{Y^0_k u_k}{1+a_{kk}Y^0_k\int_0^tu_k} 
\label{eq:Yalpha}
\end{eqnarray}

\noindent
where the auxiliary functions $u_k$ are given by

\begin{equation}
u_k(t)= \frac{ E_k(t)}{\defprod ( 1 + a_{jj} y^0_j \int_0^t u_j)^{a_{kj}/a_{jj}} }
\end{equation}
and
\beq
E_k(t)=\prod \limits_{i=1}^3(1+b_i\alpha_i^0t)^{\frac{c_{ki}}{b_i}}
\label{eq:Ek}
\eeq
Specifying to the model under consideration one finds
(see Appendix A for tabulation of the coefficients)
\beq
u_t=\frac{E_t}{(1+6Y^0_b\int u_b)^{\frac{1}{6}}
(1+4Y^0_{\lambda}\int u_{\lambda})^{\frac{1}{4}}}
\label{eq:ut}
\eeq

\beq
u_b=\frac{E_b}{(1+6Y^0_t\int u_t)^{\frac{1}{6}}
(1+4Y^0_{\tau}\int u_{\tau})^{\frac{1}{4}}
(1+4Y^0_{\lambda}\int u_{\lambda})^{\frac{1}{4}}}
\eeq

\beq
u_{\tau}=\frac{E_{\tau}}{(1+6Y^0_b\int u_b)^{\frac{1}{2}}
(1+4Y^0_{\lambda}\int u_{\lambda})^{\frac{1}{4}}}
\eeq

\beq
u_{\lambda}=\frac{E_{\lambda}}{(1+6Y^0_t\int u_t)^{\frac{1}{2}}
(1+6Y^0_b\int u_b)^{\frac{1}{2}}
(1+4Y^0_{\tau}\int u_{\tau})^{\frac{1}{4}}
(1+6Y^0_{\kappa}\int u_{\kappa})^{\frac{2}{3}}}
\label{eq:ulambda}
\eeq

\beq
u_{\kappa}=\frac{E_{\kappa}}{
(1+4Y^0_{\lambda}\int u_{\lambda})^{\frac{3}{2}}}
\label{eq:ukappa}
\eeq

We also use the shorthand notation $\int$ to mean $\int_0^t$ and drop out
for simplicity any explicit reference to the scale $t$ in all running 
quantities.
\noindent
Let us stress that \eq{Yalpha}  give the exact solution 
to $Y_k$ and $\alpha_i$ while the
 $u_k$'s given  in \eq{ut}-\eq{ukappa}, although solved formally in terms of 
the $E_k$'s and
$Y^0_k$'s as continued integrated fractions, should, in practice, be solved
iteratively. Yet, the important gain here is threefold:
 (i) as shown in \cite{Auberson}, the convergence of the
successive iterations to the exact solution can be fully
controlled analytically in terms of the initial values of the
Yukawa couplings. This means that one can in practice use truncated
iterations of Eqs.\eq{ut}-\eq{ukappa}, say to first order, and
thus obtain very good analytical approximations to the exact
solutions (ii) the structure of the solutions is explicit enough to 
allow a thorough study of some limiting regimes as we will see in the
next section (iii) furthermore, all these nice features will be naturally 
passed to the solutions for the soft SUSY breaking parameters since the latter 
will be obtained from \eq{Yalpha} through the method of \cite{Kazakov}.

\noindent
To obtain the solutions for the soft parameters one starts from
those of the couplings of the supersymmetric rigid theory,
and makes the substitutions 

\begin{eqnarray}
 \alpha & \to &\alpha_i(1+M_i\eta + \bar{M}_i
\bar\eta + 2 M_i\bar{M}_i\eta\bar\eta), \label{al} \nonumber \\
 Y_k   & \to & Y_k(1-A_k\eta -\bar{A}_k\bar \eta
+(\Sigma_k+A_k\bar{A}_k)\eta \bar\eta ), \label{yu}
\label{eq:grassmann}
\end{eqnarray}
where the $M_i$'s are the gaugino masses, the $A_k$'s the scalar soft trilinear
coupling constants and $\Sigma_k$ are the following combinations of the soft
masses 

\begin{eqnarray}
&&\Sigma_t= {m}_{Q}^2 + {m}_{T}^2  +
m^2_{2}, \;
\Sigma_b= {m}_{Q}^2 + {m}_{B}^2  +
m^2_{1}, \; 
 \Sigma_\tau= {m}_{L}^2 + {m}_{E}^2 +
m^2_{1},  \nonumber \\
&&  \Sigma_\lambda= {m}_{1}^2 + {m}_{2}^2 +
m^2_{S}, \;
 \Sigma_\kappa= 3{m}_{S}^2 
\label{eq:defSigma}
\end{eqnarray} 
and $\eta=\theta^2$ and $\bar \eta=\bar \theta^2$
are the spurion  fields depending on the Grassmannian parameters
$\theta_\alpha, \bar\theta_{\dot{\alpha}}$ (${\small \alpha} = 1,2$).

Performing the substitution \eq{grassmann} in
\eq{Yalpha}  and identifying the coefficients of the resulting polynomial
in $\eta$ and $\bar \eta$, 
the linear term in $\eta$  gives us the solution for $M_i$ and
$A_k$ and the $\eta \bar \eta$ terms the ones for $\Sigma_k$. (For
simplicity, we do not consider here CP-violating effects and take
all the soft parameters to be real valued.) The resulting exact
solutions look similar to those for the rigid couplings

\begin{eqnarray}
M_i&=& \frac{M_i^0}{1+b_i\alpha_i^0t}, \label{m} \nonumber  \\
 A_k &=& -e_k + \frac{A_k^0/Y_k^0 +a_{kk}\int
u_ke_k}{1/Y_k^0 +a_{kk}\int u_k},\label{A} \label{eq:S} \\ \Sigma_k &=&
\xi_k+A_k^2+2e_kA_k -\frac{(A_k^0)^2/Y_k^0
-\Sigma_k^0/Y_k^0+a_{kk}\int u_k\xi_k}{1/Y_k^0+a_{kk}\int u_k}, \nonumber 
\end{eqnarray}
where the new auxiliary functions $e_k$ and $\xi_k$ have been introduced
and are given in Appendix B, Eqs.\eq{et} - \eq{ekappa}.

In the particular case where $Y_b=Y_\tau=Y_{\lambda}=Y_{\kappa} =0$
Eqs.\eq{Yalpha}--\eq{S} reduce to the exact well known solutions
in the ``small $\tan \beta$" regime.

\subsection{Large Yukawa regimes}
In this section, we study various regimes of large Yukawa couplings,
and their incidence on the solutions written above.
In \cite{Moultaka}, iterated and analytical 
solutions to the RGE of the MSSM allowed an 
extensive study in the Infra-Red-Quasi-Fixed Point (IRQFP) \cite{H}
regime, i.e., 
a regime where $Y^0_{i=t,b,\tau}$ go to infinity.
In the case of the (M+1)SSM, we want to see where and
how strong could be the influence of the singlet on the MSSM solutions.
For that, we have studied 4 different regimes including the two new 
Yukawa couplings $Y^0_{\lambda}$ and $Y^0_{\kappa}$, always considering
$Y^0_{i=t,b,\tau}\rightarrow \infty$ in order to compare directly
with the MSSM case \cite{Moultaka}:
\begin{itemize}
\item{regime 1 :} $Y^0_{\lambda}$ and $Y^0_{\kappa}$ are finite
\item{regime 2 :} $Y^0_{\lambda}$ is finite and $Y^0_{\kappa}$ goes to infinity 
\item{regime 3 :} $Y^0_{\lambda}$ goes to infinity and $Y^0_{\kappa}$ is finite 
\item{regime 4 :} $Y^0_{\lambda}$ and $Y^0_{\kappa}$ go to infinity 
\end{itemize}

\noindent
Let us make here some comments about the meaning of ``infinite'' initial
conditions. Formally this means that we reach a Landau pole, and that
the corresponding low energy values are at the edge of the triviality bounds.
This clearly implies that perturbativity breaks down somewhere between
the low and high (presumably GUT) scales. 
In practice though, we have checked numerically (see section 4) 
that we reach the effective fixed point (EFP) behaviour very quickly. 
A value of $Y_i^0$ of 0.1 is already
in the EFP regime. So, there is no problem concerning the perturbativity
at high scale. (In terms of the Yukawa couplings of the Lagrangian Eqs.\eq{supot} -- \eq{scalPot} 
we never take initial values larger than 5 which corresponds
to $Y's \lsim 0.16$).

When some of the Yukawa couplings become large, one is tempted 
to drop altogether the $1$'s in
the corresponding $(1+Y^0_i \int u_i)$'s appearing in \eq{Yalpha} and 
\eq{ut}-\eq{ukappa} and to expect a typical limiting behaviour for
the $u$'s of the form\footnote{note also that this approximation 
does not work at scales too close to the initial (GUT) scale because of the 
term $\int_0^t u_i$, but it does hold for a typical electroweak scale 
$t\sim 66$.} 

\begin{equation}
u_i^{\infty} \sim \frac{1}{(Y^0)^{p_i}} \label{eq:uFP0}
\end{equation}

\noindent 
where $Y^0$ is a large Yukawa coupling.

 However, the situation is not so simple
because of the implicit dependence of each $u_i$ on the full set of $u$'s 
in a continued fraction like way, especially when not all the Yukawa couplings
become simultaneously large, as is the case in some of the regimes we consider 
here. Although it is easy to understand intuitively the validity of \eq{uFP0} 
if a similar form is obtained at some $n^{th}$ order iteration of the 
truncated approximation to Eqs.\eq{ut}-\eq{ukappa}, 
{\sl and provided that the $ p_i$'s verify $0<p_i<1$ order by order},
a more careful study is required to control the magnitudes of these powers.

Indeed, in contrast to the MSSM case \cite{Moultaka}, some of the $ p_i$'s 
can be larger than one due to the singlet sector Eqs.\eq{ulambda}, 
\eq{ukappa}. Technically, one has to solve \eq{matrice} of Appendix C
to which the reader is referred for more technical discussions and details of 
the derivation.          
%We should also point out here that, naively looking at \eq{ut}-\eq{ukappa},
%a FP limit regime for the $u_k$'s will be on the form 
%$\sim 1/(Y^0_k)^{q_k}$, for the first iteration ($q_k$ depending
%on the regime, see after). 

Here we give directly the final results for the running Yukawa couplings
in the various IRQFP regimes:
%The exact resolution is given in Appendix B, leading
%to the soft terms from the $u_k$'s
%\footnote{provided we have proven the convergence of the method as in
% \cite{Auberson}.    

%\noindent
%Intuitively, we can understand why, in a regime
%where $Y^0_i \to \infty$ and $0<p_i<1$, we can neglect $1$ in
%$(1+Y^0_i \int u_i)$, and have forms like \eq{uFP}.
%To have the complete 
%dependence, we must solve \eq{matrice}. Moreover, this approximation 
%does not work at any scale because of the term $\int_0^t u_i$, but
%we have checked that, for $t=66$, the approximation is valid.

\beq
Y_{i=t,b,\tau}^{\mathrm{FP}}=
\frac{u^{\mathrm{FP}}_{i}}{a_{ii}\int u^{\mathrm{FP}}_{t,b,\tau}}
\mathrm{~regimes~}1, 2, 3  \mathrm{~and~} 4 
\label{eq:FPtbtau} 
\eeq

\beq
Y_{\lambda}^{\mathrm{FP}}=\left\{
\begin{array}{ll}
\frac{Y^0_{\lambda}u_{\lambda}^{\mathrm{FP}}}
{(Y^0_t)^{p_{\lambda}}} \sim 0
&
\mathrm{regimes~}1, 2 \mathrm{~and~} 4 \cr \cr
\frac{u_{\lambda}^{\mathrm{FP}}}
{4 \int u_{\lambda}^{\mathrm{FP}}} 
&
\mathrm{regime~}3  \cr
\end{array} 
\right. 
\label{eq:FPlambda}
\eeq

\beq
Y_{\kappa}^{\mathrm{FP}}=\left\{
\begin{array}{lll}
\frac{Y^0_{\kappa}}{1+ 6 Y^0_{\kappa}t}
&
\mathrm{regime~}1 \cr \cr
\frac{1}{6 t}
&
\mathrm{regimes~}2 \mathrm{~and~} 4  \cr \cr
\frac{Y^0_{\kappa} u_{\kappa}^{\mathrm{FP}}}{(Y^0_t)^{\frac{3}{31}}} 
\sim 0
&
\mathrm{regime~}3  \cr \cr
\end{array} 
\right. 
\label{eq:FPkappa}
\eeq

where the $u_i^{\mathrm{FP}}$ are related to $u_i^{\infty}$ through 
\beq
u_i^{\infty}=\frac{u_i^{\mathrm{FP}}}{(Y^0_t)^{p_i}}
\label{eq:uFP}
\eeq 
and depend, as well as the $p_i$'s, on the regime under consideration
(see Appendix C).

Let us stress  several  points here :

(i) the solutions for the $Y_{i=t,b,\tau}^{\mathrm{FP}}$ have the same form 
as in the MSSM \cite{Moultaka}. Nevertheless, the effect of the 
singlet is implicit in the recursive solutions for the $u_i^{\mathrm{FP}}$
Eqs.\eq{utFP} -- \eq{ukappaFP}.

(ii) $Y^{\mathrm{FP}}_{\lambda} \sim 0$ in the regimes 1, 2 and 4. This 
will be important for the electroweak symmetry breaking discussed later on.

(iii) We have an {\it exact} analytical solution for $Y_{\kappa}$ in
the regimes 1, 2 and 4, as a function of its initial value at the GUT scale, 
$Y^0_{\kappa}$ (see the numerical analysis for a more detailed discussion).

(iv) It is also important to emphasize that the $u^{\mathrm{FP}}$ 
Eqs.\eq{utFP} -- \eq{ukappaFP} depend only on the ratios of the large initial 
values of the Yukawa couplings. However, even this dependence drops out completely in  
Eqs.\eq{FPtbtau} -- \eq{FPkappa}, so that the initial conditions
are completely screened in the IRQFP regimes as expected. Only a sensitivity
to the initial Yukawa couplings that are not large may still occur,
like in regime 1 for $Y_\kappa^{\mathrm{FP}}$, Eq.\eq{FPkappa}. 

(v) A comment is in order here about the difference between the IR quasi fixed
points we discuss and the exact fixed points
studied in \cite{Binet}. For one thing, the latter exact fixed points, 
actually exact fixed ratios, exist only in reduced couplings configurations
where all gauge couplings and all Yukawa couplings but the top are neglected,
while the quasi-fixed points we study are valid without this approximation.
Furthermore, the IRQFP's are more likely to influence the evolution
from the GUT scale to the electroweak scale than are the exact fixed points
\cite{H}. Let us note, however, that in one or the other of our four
IRQFP regimes we find either $\kappa$ or $\lambda$ to be vanishingly
small, Eqs.\eq{FPlambda}, \eq{FPkappa}, similarly to the case of two
among the three exact fixed point regimes determined in \cite{Binet}. 
Nonetheless, the latter two regimes were found to be infrared repulsive 
\cite{Binet}, while as one can infer from the structure of the denominator in 
Eq.\eq{Yalpha}, the top down evolution of the Yukawa couplings tends generically 
always {\sl towards} the IRQFP's behaviour.\\

\noindent
To find the IRQFP behaviour of the soft parameters $A_i, \Sigma_i$
one can either perform the substitutions Eq.\eq{grassmann}  
in Eqs.\eq{FPtbtau} -- \eq{FPkappa}, \eq{utFP} -- \eq{ukappaFP}, \eq{Ek},
or study the large initial Yukawa limit directly from the general
solutions Eq.\eq{S}. Denoting the auxiliary functions $e_i$ 
of Eqs.\eq{et} - \eq{ekappa}  by $e^{\infty}_i$ in this limit, 
we obtain for the $A$'s

\beq
A_{i=t,b,\tau}^{\infty}=
-e_i^{\infty}+\frac{\int u_i^{\infty}
e_i^{\infty}}
{\int u_i^{\infty}}
 = A_i^{\mathrm{FP}} \mathrm{~in~all~regimes}
\label{eq:AFPtbtau}
\eeq

\beq
A_{\lambda}^{\infty}=\left\{
\begin{array}{ll}
-e_{\lambda}^{\infty}+A^0_{\lambda}
&
\mathrm{regimes~}1, 2  \mathrm{~and~} 4\cr \cr
-e_{\lambda}^{\mathrm{FP}}+
\frac{\int u_{\lambda}^{\mathrm{FP}}e_{\lambda}^{\mathrm{FP}}}
{\int u_{\lambda}^{\mathrm{FP}}}
&
\mathrm{regime~}3  \cr \cr
\end{array} 
\right. 
\label{eq:AFPlambda}
\eeq

\beq
A_{\kappa}^{\infty}=\left\{
\begin{array}{lll}
-e_{\kappa}^{\infty}+
\frac{A^0_{\kappa}+6 Y^0_{\kappa} \int e_{\kappa}^{\infty}}
{1+6 Y^0_k  t}
&
\mathrm{regime~}1 \cr \cr
-e_{\kappa}^{\mathrm{FP}}+\frac{\int e_{\kappa}^{\mathrm{FP}}}{t}
&
\mathrm{regimes~}2 \mathrm{~and~} 4  \cr \cr
-e_{\kappa}^{\infty}+A^0_{\kappa}
&
\mathrm{regime~}3  \cr \cr
\end{array} 
\right.
\label{eq:AFPkappa}
\eeq

\noindent
where 

\begin{equation}
A_i^{\mathrm{FP}} \equiv -e_i^{\mathrm{FP}}+\frac{\int u_i^{\mathrm{FP}}
e_i^{\mathrm{FP}}}
{\int u_i^{\mathrm{FP}}} \label{eq:defAFP}
\end{equation}

\noindent
and the $e^{\infty}_i$ are defined in Appendix D Eqs.\eq{einftyt} -- 
\eq{einftykappa}. These auxiliary functions depend explicitly on the initial 
values $A^0_k$. In some IRQFP regimes, this dependence cancels out
in the running parameters $A_i^{\infty}$.
Such an independence occurs in all the regimes for $A_{t,b,\tau}$ 
Eq.\eq{AFPtbtau} as was the case in the MSSM \cite{Moultaka}, 
in the regime 3 for $A_\lambda$ Eq.\eq{AFPlambda} and in regimes 2, 3 for 
$A_\kappa$ Eq.\eq{AFPkappa}. 
In such cases we have re-expressed the results in terms of new quantities 
$e^{\mathrm{FP}}_i$ which are independent of the initial conditions.
The dependence on initial conditions is explicited further in 
Eqs.\eq{Ainftylambda}, \eq{Ainftykappa}.\\

\noindent
Similarly, denoting by $\xi^{\infty}_i$ the auxiliary functions $\xi_i$ 
Eqs.\eq{xit} - \eq{xikappa} in the large Yukawa  
limits we obtain for the $\Sigma$'s 

\beq
\Sigma_{i=t,b,\tau}^\infty=\Sigma_{i}^{\mathrm{FP}}\equiv
\xi_i^{\mathrm{FP}}+(A_i^{\mathrm{FP}})^2
+2e_i^{\mathrm{FP}}A_i^{\mathrm{FP}}
-\frac{\int u_i^{\mathrm{FP}}\xi_i^{\mathrm{FP}}}{\int u_i^{\mathrm{FP}}}
\eeq

\noindent

\beq
\Sigma_{\lambda}^\infty= 
\left\{
\begin{array}{ll}
\xi_{\lambda}^{\infty}
+(A_{\lambda}^{\infty})^2
+2A_{\lambda}^{\infty}e_{\lambda}^{\infty}
+\Sigma^0_{\lambda}-(A^0_{\lambda})^2
&
\mathrm{regime~} 1,2 \mathrm{~and~} 4 \cr \cr
\xi_{\lambda}^{\mathrm{FP}}+(A_{\lambda}^{\mathrm{FP}})^2
+2e_{\lambda}^{\mathrm{FP}}A_{\lambda}^{\mathrm{FP}}
-\frac{\int u_{\lambda}^{\mathrm{FP}}\xi_{\lambda}^{\mathrm{FP}}}
{\int u_{\lambda}^{\mathrm{FP}}}
&
\mathrm{regime~} 3
\cr 
\end{array}
\right.
\eeq

\beq
\Sigma_{\kappa}^\infty=
\left\{
\begin{array}{ll}
\xi_\kappa^{\infty}
+(A_\kappa^{\infty})^2
+2A_\kappa^{\infty} e_\kappa^{\infty}
+\xi_\kappa^{\infty}
+\frac{\Sigma^0_\kappa-(A^0_\kappa)^2-6Y^0_\kappa\int \xi_\kappa^{\infty}}
{1+6Y^0_\kappa t}
&\mathrm{regime~}1 \cr \cr
\xi_{\kappa}^{\mathrm{FP}}+(A_{\kappa}^{\mathrm{FP}})^2
+2e_{\kappa}^{\mathrm{FP}}A_{\kappa}^{\mathrm{FP}}
-\frac{\int \xi_{\kappa}^{\mathrm{FP}}}{t}
&
\mathrm{regimes~} 2 \mathrm{~and~} 4 \cr \cr
\xi_{\kappa}^{\infty}
+(A_{\kappa}^{\infty})^2
+2A_{\kappa}^{\infty}e_{\kappa}^{\infty}
+\Sigma^0_{\kappa}-(A^0_{\kappa})^2
&
\mathrm{regime~} 3
\cr 
\end{array}
\right.
\eeq

\noindent
where again $\xi_i^{\mathrm{FP}}$ have been used instead of $\xi_i^{\infty}$ 
for those cases where the dependence on initial conditions is absent
in the running $\Sigma$'s.  The reader is referred to
Appendix D for an extended discussion of the relation between
$e_i^{\mathrm{FP}}, \xi_i^{\mathrm{FP}}$ and $e_i^{\infty}, \xi_i^{\infty}$ 
and for the explicit dependence on initial conditions 
Eqs.\eq{siginftylambda}, \eq{siginftykappa} 
Some remarks are in order:

(i) $A_{\lambda}$ depends on the initial condition $A^0_i$ in the regimes
1, 2 and 4, but in these regimes $Y_{\lambda}\sim 0$ Eq.\eq{FPlambda}.
Similarly, $A_{\kappa}$ depend on the initial values of $A^0_i$ in
the regime 3, but in this regime $Y_{\kappa} \sim 0$ \eq{FPkappa}.
So, we can conclude that, in every regime, the running of the combination
$4 \pi \sqrt{Y_i} A_i$, 
$i=\lambda, \kappa$, the one present in the Lagrangian, 
is indeed screened from the GUT-scale initial conditions.

(ii) We have the $exact$ analytical dependence of the soft terms  on 
the initial conditions $A^0_i, \Sigma_i^0$ ($c.f.$ Appendix D).  

(iii) Whereas the dependence of the $\xi^{\infty}$ on the products 
$A^0_iA^0_j$ and $\Sigma^0_j$ is rather complicated,
the $\Sigma_i^{\infty}$'s depend only on the $\Sigma^0_j$, 
except for regime 1 where $A^0_\kappa$ is present, see
Eqs.\eq{siginftylambda}, \eq{siginftykappa}.
Moreover, this dependence is exactly the same as the dependence of the
$A^{\infty}_i$ on the $A^0_j$. 

%This is due to the matricial form of the
%equations for the $\xi$ (see Appendix C). 
%The next section is devoted to this analysis.

(iv) It is worth noting that the abovementioned sensitivity
of $\Sigma^{\infty}_{\lambda, \kappa}$ to $Y^0_{\kappa}, A^0_\kappa$
 in the regime 1 disappears in all the soft masses, 
except for the singlet soft mass ($m_S$), see Eqs.\eq{mTsoft} --\eq{mSsoft},
\eq{msoft3}. Moreover, in  regimes 1, 2 and 4, where the soft mass of the 
singlet separates from all the others,  Eq.\eq{msoft124}, the dependence on 
initial conditions for the soft masses is exactly the same as the one found in
 \cite{Moultaka} in the case of the MSSM.

\noindent
Finally, let us note that it does not seem possible to give the explicit 
dependence of the $A$ and the $\Sigma$'s on gauginos mass initial conditions
because the latter come always in scale dependent contributions,
Eqs.\eq{dtildeE}, \eq{d2tildeE}.\\

\noindent
A further point should be made here about the evolution of the
soft scalar masses since we address the most general situation
beyond universality.   
In fact, to solve exactly the RGE for the soft masses, we have
to consider the complete equations, including a $U(1)$ induced
 ``trace term'' $S$,

\beq
\dot{(m^2_i)}=f_i(M_1,M_2,M_3,\Sigma_i,A_i)+T_i\alpha_1 S 
\label{eq:RGEm}
\eeq 

\noindent
where $f_i(M_{1,2,3},\Sigma_i,A_i)$ are defined as

\bea
f_t&=&\frac{16}{9}\alpha_1 M_1^2+\frac{16}{3} \alpha_3 M_3^2
-2Y_t(\Sigma_t+A_t^2)
\nonumber
\\
f_b&=&\frac{4}{9} \alpha_1 M_1^2+\frac{16}{3} \alpha_3 M_3^2 
-2Y_b(\Sigma_b+A_b^2)
\nonumber
\\
f_Q&=&\frac{1}{9} \alpha_1 M_1^2 +3 \alpha_2 M_2^2+\frac{16}{3} \alpha_3 M_3^2
-Y_t(\Sigma_t+A_t^2)-Y_b(\Sigma_b+A_b^2)
\nonumber
\\
f_{E}&=&4 \alpha_1M_1^2-2Y_{\tau}(\Sigma_{\tau}+A_{\tau}^2)
\nonumber
\\
f_{L}&=&\alpha_1 M_1^2+3\alpha_2 M_2^2
-Y_{\tau}(\Sigma_{\tau}+A_{\tau}^2)
\nonumber
\\
f_{H_1}&=&\alpha_1 M_1^2 + 3 \alpha_2 M_2^2
-Y_{\lambda}(\Sigma_{\lambda}+A_{\lambda}^2)
-Y_{\tau}(\Sigma_{\tau}+A_{\tau}^2)-3Y_b(\Sigma_b+A_b^2)
\nonumber
\\
f_{H_2}&=&\alpha_1 M_1^2+ 3 \alpha_2 M_2^2
-Y_{\lambda}(\Sigma_{\lambda}+A_{\lambda}^2)
-3Y_t(\Sigma_t+A_t^2)
\nonumber
\\
f_s&=&-2Y_{\lambda}(\Sigma_{\lambda}+A_{\lambda}^2)
-2Y_{\kappa}(\Sigma_{\kappa}+A_{\kappa}^2) \label{eq:deff}
\eea 

\beq
S=\sum_{generations} (m_{Q}^2-2m_{U}^2+m_{D}^2-m_L^2+m_{E}^2)+m_{2}^2-m_{1}^2
\label{eq:Sdef}
\eeq
and  
\beq
T_i=
\{
1/3,-4/3,2/3,-1,2,-1,1,0
\} \label{eq:coefs}
\eeq
with $i=\{ Q,U,D,L,E,H_1,H_2,S \}$ respectively. From Eqs.\eq{RGEm}, \eq{deff},
\eq{Sdef} and \eq{coefs} one sees that
$\dot{S} \propto \alpha_1 S$, so that if $S$ vanishes at some scale $t_0$
it will vanish identically at any scale $t$. For instance $S=0$ at any scale in 
the case of universality of all soft scalar masses. More 
generally, $S$ can still be ignored even when universality is relaxed provided
that the initial conditions are such that $S(t_0)=0$. This simplifying
configuration was taken up in \cite{Moultaka}.
Generically, however, one should solve Eq.\eq{RGEm} keeping the trace term.
This can be easily done by writing the solution for the soft masses as 

\beq
m_i^2=(m_i^2)_{f}+(m_i^2)_{Tr} \label{eq:solm}
\eeq 

\noindent
where $(m_i^2)_{f}$ is solution of the equation
$\dot{(m^2_i)}_f=f_i(M_1,M_2,M_3,\Sigma_i,A_i)$, 
and $\dot{(m_i^2)}_{Tr}=T_i\alpha_1 S$.
The $(m^2_i)_f$ are then given by Eqs.\eq{mTsoft}, \eq{mSsoft}, 
and $(m_i^2)_{Tr}$ is just the solution of a linear differential
 equation system.
We obtain

\beq
(m_i^2)_{Tr}= t_i S_0 ( (1+b_1\alpha_1^0t)^{\frac{26}{33}}-1 )
\label{eq:mtrace}
\eeq 

%\noindent
%with 
%\beq
%T=(1+b_1\alpha_1^0t)^{\frac{26}{33}}-1
%\eeq

\noindent
with

\beq
t_i=\frac{1}{26}
\{
1,-4,2,-3,6,-3,3,0
\}
\eeq

\noindent
where $S_0 = S(t=0)$.

\section{Numerical analysis}
\renewcommand{\theequation}{4.\arabic{equation}}
\setcounter{equation}{0}

As we stressed before, one of the advantages of our approach
is that we can extract the exact sensitivity to initial
conditions in the quasi-fixed point regimes.
 In this section, we study numerically the solutions
for the Yukawa couplings $Y_i$, the
trilinear couplings $A_i$, and the soft parameters $m_i$, in
the four IRQFP regimes we have identified in section 3. We
compare  our analytical solutions to the results obtained
by a purely numerical resolution of the RGE's, relying on a FORTRAN
code which evolves the relevant parameters from the GUT
scale to the electroweak scale through an algorithm similar to the one
used in \cite{ERS2}.

\noindent
In the case of the Yukawa couplings we will restrain ourselves,
for the sake of illustration,  to the study 
of $Y_\kappa$. In Fig. 1a we plot $Y_{\kappa}$ at the EW scale $(t \simeq 66)$
as a function of its value at the GUT scale, for different
values of $Y^0_{\lambda}$  and for $Y^0_{t,b, \tau} =0.1$.
Since $Y_{\kappa}^0$ is varied only in a range of small values, 
we reach in the same graph regime 1 
(finite $Y^0_{\lambda}$ and $Y^0_{\kappa}$) as well as regime 3  
(infinite $Y^0_{\lambda}$ and finite $Y^0_{\kappa}$)
\footnote{As far as the numerics are concerned, finite $Y^0_i$ means 
$Y^0_i \sim 10^{-2}$ and infinite $Y^0_i$ means $Y^0_i \sim 10^{-1}$.}. 
We have also plotted in the same figure the analytical solution
\eq{FPkappa}. We thus see, in the small $Y_{\lambda}^0$ region,
 the numerical agreement with the  the expected
behaviour  $Y_{\kappa}(t=66) \sim \frac{Y^0_{\kappa}}{1+6Y^0_{\kappa}t}$
of regime 1, and $Y_{\kappa}(66) \sim 0$
for large values of $Y^0_{\lambda}$ in accord with regime 3.
Fig. 1b shows the variation of $Y_{\kappa}(t=66)$ 
for bigger values of $Y^0_{\kappa}$ (regime 2 and 4).
We clearly see that $Y_{\kappa}$ tends to the analytical 
solution $\frac{1}{a_{\kappa \kappa}t}$ ($\sim 0.0025$ for $t=66$)
for small values (regime 2) or high values (regime 4) of 
$Y^0_{\lambda}$. Evaluating the $u^{\mathrm{FP}}_i$'s
up to the third order iteration with a procedure similar to that
in \cite{Moultaka}, we also found very good numerical agreement between
\eq{FPtbtau}, \eq{FPlambda} and the FORTRAN code output.\\

\noindent
We note finally that  our results for $Y_{\kappa}$ in regimes 2 and 4
are in perfect agreement with the numerical illustrations of \cite{Ellis},
while $Y_{\lambda}$ was found to differ slightly. The reason for this
numerical difference can be traced back to the fact that $Y_{\kappa}$ 
is completely decoupled from the $t, b, \tau$ sector, while 
$Y_{\lambda}$ is closely tied to it.
Whence an important influence of the $Y_{t, b, \tau}$ on the
running of $Y_{\lambda}$, which modifies significantly its low scale value.

%\vskip 4cm
%\begin{figure}[h]

%\newpage
\noindent
\vskip 2.5cm
%\hskip -2cm
\input fig1.fignew
\input fig2.fignew
\vskip 0.5cm
\hskip 3.2cm {\bf Fig. 1.a} \hskip 7.5cm {\bf Fig. 1.b}

\noindent
In figures 2a, 2b we plot the coefficients of the various initial conditions
$A^0_i,  (i=t,b, \tau, \lambda, \kappa)$
 at the GUT scale  which enter
the running $A_{\lambda}$ at the electroweak scale $(t=66)$ as a function
of $Y^0_{\lambda}$.  
Since $Y^0_{\lambda}$ is varied in a large range of values and the illustrations
are given for one very small and one very large value of $Y^0_{\kappa}$,
Fig. 2a, 2b cover all four regimes. Furthermore the initial conditions
for $Y^0_{i=t,b,\tau}$ are fixed to the (large) common value $0.1$.
Thus when $Y^0_{\lambda}$ runs from 0 to 5 one observes the transition
from regime 1 to regime 3 in Fig. 2a and from regime 2 to regime 4
in Fig. 2b. [We should keep in mind that the very large values taken by 
$Y_\lambda^0$ or $Y_\kappa^0$ are only for the sake of numerical comparison
in the deep IRQFP regimes at the electroweak scale. At much higher
scales they eventually lead to perturbatively non reliable results.]
The trend of the plots is in perfect agreement with what is anticipated
from Eq.\eq{Ainftylambda}. In Fig. 2a one retrieves the coefficients
of regime 1 in the region of small $Y^0_\lambda$. For large $Y^0_\lambda$
all the coefficients vanish asymptotically as expected in
regime 3. [The asterisks ({\bf*}) at the right of the graph 
represent the values of the coefficients for $Y^0_{\lambda}=10$.]  
In Fig 2b where $Y^0_{\kappa}$ is taken very large $(=20)$  
we observe the evolution as a function of $Y^0_{\lambda}$ from the regime 2, 
to the regime 4. The weaker sensitivity to $Y^0_{\lambda}$  (as compared to
Fig. 2a) corroborates here the fact that the two regimes 2, 4 lead to 
identical coefficients in this case, see Eq.\eq{Ainftylambda}.

%Moreover, if we look in more detail to the expression (D.2) and (D.3), we
%can find in the hypothesis of \cite{ERS2}
%\footnote{{\it i.e.} Universality at the GUT scale, and 
%neglecting all matter Yukawa, except the {\it top} one}
% the same approximations
%for the trilinear couplings and masses. Except that, in our case, we have the 
%{\it exact} and {\it analytical} dependence on the initial values
%$A_i^0$ and $m_i^0$. We can checked that they are in the condition of our 
%regime 1. See the Electroweak Symmetry Breaking for a more detailed discussion.

%\newpage
%\noindent
%\hskip -2cm
%\input fig1.fig
%\input fig2.fig
%\vskip 3cm
\noindent
\vskip 0.5cm
%\hskip -2cm
\input fig3.fignew
\input fig4.fignew
\vskip 0.5cm
\hskip 3.2cm {\bf Fig. 2.a} \hskip 7.5cm {\bf Fig. 2.b}

\noindent
We have made the same analysis for the soft terms $\Sigma_i$ 
and $m_i$ and obtained numerical results in total agreement with
our analytical expression (Appendix D). We just present here
in Fig. 3a, 3b, respectively 
the dependence of $(m_Q)^2$ and $(m_1)^2$ on the initial GUT scale
values of all soft squared masses (not including the trace contribution
of Eq.\eq{mtrace}, for a small value of $Y^0_{\kappa}$ (0.001), and for
$Y^0_{\lambda}$ running from 0 to 5. One sees the transition
from regime 1 to regime 3 with increasing $Y^0_{\lambda}$.
The dependence on the soft mass $(m^0_S)^2$ vanishes towards
regime 1.

%\vskip -0.5cm
%\hskip -2cm
\input fig5.fignew
\input fig6.fignew
\vskip 0.5cm
\hskip 3.2cm {\bf Fig. 3.a} \hskip 7.5cm {\bf Fig. 3.b}

%\vspace{-0.5cm}

\noindent
Finally we illustrate in Fig.4 the behaviour of $Y_\lambda, Y_\kappa$ in
all four regimes, to stress the fact that these regimes do set in effectively, 
well before that the initial conditions $Y_\lambda^0, Y_\kappa^0$ become
infinitely large.  
%\begin{center}

\hskip 4cm
\input fig1bis.fignew

\vskip 0.5cm
\hskip 5 cm {{\bf Fig. 4}: $Y_\lambda$ (full lines), $Y_\kappa$ (dotted lines) }

\section{Electroweak symmetry breaking}
\renewcommand{\theequation}{5.\arabic{equation}}
\setcounter{equation}{0}
In this section we consider the impact of the four IRQFP 
regimes on more phenomenological
issues. It turns out, as we will show hereafter,
that regimes 2 and 4 are consistent with
the requirement of EWSB only for very small
$|\hat{\mu}|$. These regimes are thus disfavored or excluded from present 
experimental exclusion lower bounds on the lightest chargino mass \cite{LEP, LEP1}.  
Regimes 1 and 3 do not suffer from such features and correspond to viable 
configurations of least sensitivity to initial conditions.\\

\noindent
The study is carried out at the  tree-level.
[Some of the conclusions will remain true if one-loop corrections
are included, but we will not discuss the issue further in the present
paper.] 
The EWSB conditions involving the three Higgs v.e.v.'s are obtained from
 Eq.\eq{scalPot}
in the form:

\begin{eqnarray}
&&h_1 \; \large [ m_1^2 + \lambda^2 (h_2^2 + s^2) +\frac12 \bar{g}^2 (h_1^2 - h_2^2) + 
(A_{\lambda} \lambda s + \kappa \lambda s^2) \frac{h_2}{h_1} \large ]= 0 
\label{ewsb1} \\
&&h_2 \; \large [ m_2^2 + \lambda^2 (h_1^2 + s^2) +\frac12 \bar{g}^2 (h_2^2 - h_1^2) + 
(A_{\lambda} \lambda s + \kappa \lambda s^2) \frac{h_1}{h_2} \large ]= 0 
 \label{ewsb2} \\
&&s \large \; [ m_s^2 + \lambda^2 (h_1^2 + h_2^2) + 2 \kappa^2 s^2 + 
A_{\kappa} \kappa s + (\frac{A_\lambda}{s} + 
2 \kappa) \lambda h_1 h_2 \large ] =0  \label{ewsb3} 
\end{eqnarray}

where 

\begin{eqnarray}
<H_1> \equiv \left ( \begin{array}{c}
               h_1 \\
                0   \\
               \end{array} \right )  \hspace{2cm}
<H_2> \equiv \left( \begin{array}{c}
                     0\\
                     h_2 \\
                     \end{array} \right ) \hspace {2cm}
<S> \equiv s
\end{eqnarray} 

Here $h_1, h_2$ and $s$ are chosen to be real-valued and positive. 
(see for instance \cite{Ellis} for a discussion of the freedom
in the choice of the various parameters). 

In addition to Eq.(\ref{ewsb1} - \ref{ewsb3}) the requirement of 
correct electroweak scale reads

\begin{equation}
h_1^2 + h_2^2 = \frac{M_Z^2}{\bar{g}^2} \label{Zmass}
\end{equation}

From equations (\ref{ewsb1}), (\ref{ewsb2}) and (\ref{Zmass}), and provided
that non of the v.e.v.'s is vanishing, one easily gets

\begin{eqnarray}
h_1^2 &=& \frac{M_Z^2}{\bar{g}^2} \frac{ \frac{1}{2} M_Z^2 + m_2^2 + \lambda^2 s^2}{ M_Z^2 +
m_1^2 + m_2^2 + 2  \lambda^2 s^2} \label{eqh1} \\
h_2^2 &=& \frac{ M_Z^2}{\bar{g}^2} \frac{ \frac{1}{2} M_Z^2 + m_1^2 + \lambda^2 s^2}{ M_Z^2 +
m_1^2 + m_2^2 + 2  \lambda^2 s^2} \label{eqh2} \\
h_1 h_2 &=& -\frac{ M_Z^2}{\bar{g}^2} 
\frac{(A_{\lambda} \lambda s + \kappa \lambda s^2)}{ m_1^2 + m_2^2 + 
  \lambda^2 (\frac{M_Z^2}{\bar{g}^2} + 2 s^2) } \label{eqh3} 
\end{eqnarray}

Note that the above equations yield the familiar EWSB conditions of the MSSM
when one goes to the limit  $\lambda, \kappa \to 0, s \to \infty$ with 
$ \lambda s, \kappa s$ finite, with the identifications 
$\hat{\mu} \equiv \lambda s, B\equiv A_{\lambda} + \kappa s$, 
$\tan \beta \equiv h_2/h_1, h_1 h_2 \equiv M_Z^2 \sin 2 \beta/\bar{g}^2$ 
(of course with the proviso of Eq.(\ref{ewsb3}) which correlates dynamically 
$s$, thus $\hat{\mu}$, to the other parameters).

From Eq.(\ref{eqh3}), one obtains\footnote{ we use from now on the shorthand
notation, $t_\beta, s_{2\beta}, c_{ 2\beta}$ for $\tan \beta, \sin 2 \beta,
\cos 2 \beta$.} 

\beq
\frac{\kappa}{\lambda}=a+b\lambda^2
\label{kslam}
\eeq

\noindent
with
\bea
a&=&-\frac{A_{\lambda}}{\hat{\mu}}
+\frac{1}{2}\frac{M_Z^2}{\hat{\mu}^2}s_{2 \beta}
(1-\frac{1/2M_Z^2+m_1^2+\hat{\mu}^2}{m_t^2}\frac{y_t^2}{\bar{g}^2})
\\
&=& -\frac{A_{\lambda}}{\hat{\mu}} -\frac{1}{ t_\beta} 
- \frac{1}{2 t_\beta \hat{\mu}^2} (c_{2 \beta} M_Z^2 + 2 m_1^2 )
\label{kslam1} \\
b&=&-\frac{1}{2}\frac{M_Z^2}{\hat{\mu}^2}\frac{s_{2 \beta}}{\bar{g}^2}
\eea
\noindent
Equation (\ref{kslam}) shows a linear correlation between $\kappa$ and $\lambda$
(in the regime of small $\lambda (\ll \bar{g}^2)$)
even at low energy, the coefficient $a$ depending on $y_t$. This was noticed
numerically in \cite{King} for $\lambda^0$ and $\kappa^0$ in constrained
cases (universality, $y_t \gg y_b$...).

On the other hand, the scalar potential at the electroweak symmetry breaking minimum is obtained 
from Eqs.\eq{scalPot}, (\ref{ewsb1}-\ref{ewsb3}) and reads

\begin{eqnarray}
V_{min} &=& - ( \kappa s + \lambda h_1 h_2)^2 - \lambda^2 s^2 (h_1^2 + h_2^2) \nonumber \\
&&  - A_{\lambda} \lambda s h_1 h_2 -\frac{1}{3} A_{\kappa} \kappa s^3 -\frac{1}{4} M_Z^2 
(h_1^2 + h_2^2) c_{2 \beta} \label{potmin}
\end{eqnarray}

Eqs.(\ref{eqh1} -- \ref{potmin}) will be very useful
when discussing the implication of IRQFP regimes.

As previously stressed, although the IRQFP regimes correspond, strictly
speaking, to some or all of the Yukawa couplings taking infinite values
at some high energy scale, in practice we stay away from these unphysical
configurations. Still, the typical IRQFP is essentially preserved as can be seen
 from Fig.4,  and gives a very good idea of the sensitivity to initial 
conditions. 

Bearing this in mind, we thus refer in the following to regime 1 as implying
$\kappa, \lambda \ll \bar{g}$, to regimes 2,4 as implying 
$\lambda \ll \kappa, \bar{g} $ and to regime 3
as $\kappa \ll \lambda,\bar{g} $ at the electroweak scale, 
rather than the more strict behavior given in 
Eqs.(\eq{FPlambda}, \eq{FPkappa}).
Since the discussion 
may get a little too involved, 
especially in regime 3, the reader interested only in
the conclusions can go directly to the summaries in {\bf 5.2} and {\bf 5.4}.

\subsection{Regimes 1, 2, 4:}

Note first that  Eqs.(\ref{eqh1}, \ref{eqh2}) do not have a solution
if $\lambda =0$ with $s$ finite, unless $m_1^2 + m_2^2$ is fine-tuned
to zero in which case there is an infinitely degenerate set (a valley) 
of solutions for $h_1, h_2$. This degeneracy is not lifted by one-loop 
corrections as can be seen for instance from the form of the top/stop 
contributions \cite{ERS1} and should thus be discarded as non physical.
We now take $\lambda$ non vanishing but small enough 
to allow a reliable expansion to first order. Moreover, we consider separately
the cases  (a) $s \lsim m, A$ and (b) $s \gg m, A$ where $m, A$ denote
generically the magnitudes of the soft masses and couplings. 
We will also denote by $M_{soft}$
a generic value of the soft masses and couplings, or the smallest of these values.\\

\begin{itemize}
\item[(a)]{ $s \lsim m, A$:}
%{\bf (a) $s \lsim m, A$:}

In this case $|\lambda| s \ll  m, A$ and an expansion in small parameter  is reliable. 
Performing this expansion in Eqs.(\ref{eqh1} -- \ref{eqh3}) to second order 
in $\hat{\mu}/M_{soft}$ (and first order in $\epsilon$) and eliminating 
$h_1, h_2$, the equation determining dynamically $\hat{\mu}( \equiv \lambda s)$ takes the simple form:

\begin{equation}
\zeta \hat{\mu}^2 + \frac{A_\lambda}{m_1^2 + m_2^2} ( 1 - \epsilon) \hat{\mu} + 
\frac{\sqrt{ (\frac{1}{2} M_Z^2 + m_1^2)(\frac{1}{2} M_Z^2 + m_2^2)}}{| M_Z^2 + m_1^2 + m_2^2 |} + 
O((\frac{\hat{\mu}}{M_{soft}})^3)= 0 \label{mureg124}
\end{equation}

where 
\begin{eqnarray}
\epsilon & = & \frac{\lambda^2}{\bar{g}^2} \frac{M_Z^2}{m_1^2 + m_2^2} \label{eps} \\
\zeta &= &  
\frac{ (m_1^2 - m_2^2)^2}{\sqrt{ (M_Z^2 + 2 m_1^2)( M_Z^2 + 2 m_2^2)} | M_Z^2 + m_1^2 + m_2^2 | 
( M_Z^2 + m_1^2 + m_2^2 )} \nonumber \\
&&+ \frac{1}{(m_1^2 + m_2^2)} \frac{\kappa}{\lambda} ( 1 - \epsilon) \label{zet} 
\end{eqnarray}

\underline{{\sl regimes 2, 4:}}

\noindent
Here $|\lambda| \ll |\kappa|, \bar{g}$. Solving Eq.(\ref{mureg124}) for $\hat{\mu}$ leads in this case to

\begin{equation}
\hat{\mu} = \pm \sqrt{-\frac{\lambda}{\kappa} ( m_1^2 + m_2^2)} \sqrt{\frac{ 
\sqrt{ (\frac{1}{2} M_Z^2 + m_1^2)(\frac{1}{2} M_Z^2 + m_2^2)}}{| M_Z^2 + m_1^2 + m_2^2 |}}
+ O(\frac{\lambda}{\kappa}) \label{reg24eq1}
\end{equation}

which, apart from the consistency conditions $(\frac{1}{2} M_Z^2 + m_1^2)(\frac{1}{2}M_Z^2 + m_2^2) \geq 0$ and  
$ \lambda \kappa  ( m_1^2 + m_2^2) \leq 0$,
shows that $|\hat{\mu}|$ tends to be generically very small being suppressed
by the size of $\sqrt{|\lambda/\kappa|}$. Thus, even if one chooses the soft
parameters sufficiently larger than the electroweak scale, so that the 
condition $|\lambda| s \ll  m, A$ dictated by regimes 1, 2, 4 does not imply 
{\sl a priori} small $|\hat{\mu}|$ in comparison to the electroweak scale, then
the dynamics of regimes 2, 4 will still lead to a vanishing $\hat{\mu}$.
A vanishing $\hat{\mu}$ implies a lightest chargino to be lighter
than $M_W$ and drops even much lower, for $\tan \beta >1$, 
see Eq.(\ref{supbound}). Such  a configuration would be excluded, or at most
marginally acceptable, by the LEPII lower bounds on the lightest chargino
\cite{LEP, LEP1}, were it not for the fact that, since $\lambda$ is small in 
the regimes under consideration, unconventional signatures due to displaced 
vertices can emerge in the case of the (M+1)SSM thus requiring a reanalysis of
the experimental data \cite{EH2}. Even so, the theoretical upper bound 
(\ref{supbound}) with vanishing $\hat{\mu}$ will start conflicting with the 
conservative LEPI kinematical limit of $M_Z/2$ as soon as $\tan \beta > 2.27$ 
or so.    
 
\underline{ {\sl regime 1:}} 

\noindent
This regime has a crucial difference with 2 and 4, namely that $\kappa$ and $\lambda$ can be generally 
of the same order. In this case the behavior given in Eq.(\ref{reg24eq1}) is
no more valid. One can of course still solve readily Eq.(\ref{mureg124})
keeping in mind that $|\hat{\mu}| \ll m, A$ in the regime under
consideration.    
Since the $\hat{\mu}$ independent term in Eq.(\ref{mureg124}) is generically of 
order $1/2$ it is straightforward to see, taking into account 
Eqs.(\ref{eps}, \ref{zet}) that a consistent
$\hat{\mu}$ requires that $A_{\lambda}$ be very large compared to 
$\sqrt{ m_1^2 + m_2^2}$, at the
electroweak scale. This suggests that,  
generically ({\sl i.e} discarding fine-tuned cancellations
in $m_1^2 + m_2^2$), regime 1 disfavors solutions with $s \lsim m, A$
if the relevant soft parameters are of the same order of magnitude
at some initial scale.  
%(ex. $A_0 \sim m_0$ which would be natural in SUGRA motivated universality 
%assumptions).  

\item[(b)]{ $s \gg m, A$:}
%{\bf (b) $s \gg m, A$:}

In this case $|\lambda| s \gsim m , A$.

\underline{{\sl regimes 2,4:}} 

\noindent
Eq.(\ref{eqh3}) takes the form

\begin{equation}
h_1 h_2 =  - \frac{\kappa}{\lambda} \hat{\mu}^2 \frac{ M_Z^2}{\bar{g}^2} \frac{1}{(m_1^2 +m_2^2 + 2 \hat{\mu}^2)^2} + O(\lambda^0)
\end{equation}

that is, $h_1 h_2$ can be made arbitrarily large in the deep 2,4 regions ( $\frac{\lambda}{\kappa} \to 0$ ).
This is however in contradiction with the behavior dictated by Eqs.(\ref{eqh1}, \ref{eqh2}) in the same
regions and leads to an inconsistency, even if  $M_Z^2 + m_1^2 + m_2^2 + 2 \hat{\mu}^2$ is artificially 
fine-tuned to $O(\lambda)$.

\underline{{\sl regime 1:}}
 
%\vspace{-0.2cm}
\noindent
Here no simple expressions can be derived and one would have to solve the 
full-fledged higher order polynomial equation for $\hat{\mu}$ combined from  
Eqs.(\ref{eqh1}-- \ref{eqh3}).

\end{itemize}

\subsection{ Summary for regimes 1, 2, 4}
We have shown that the IRQFP regimes 2,4 can be consistent with 
EWSB only  when $s$ is of the order of the soft masses, in which case 
$|\hat{\mu}|$ becomes too small to be consistent with present limits on 
chargino masses (or at best marginally consistent if $\tan \beta \lsim 2.3$
when only conservative LEPI limits are considered). 
On the other hand we found that for dynamical reasons, and independently of 
any phenomenological considerations, regime 1
 can be consistent only for large $s$ ($\gg m, A$) . This last point is one 
ingredient for the explanation of the 
numerically established large values of 
singlet Higgs v.e.v., \cite{ERS1}. We will come back later to these features. 
Let us also note that the above conclusions were drawn without taking
into account Eq.(\ref{ewsb3}). This equation can be viewed in this context as 
nearly correlating the two extra soft parameters $m_s, A_\kappa$ and enters 
the game as a further constraint translated to initial conditions at some high 
energy scale.   

\subsection{Regime 3:}

Here $|\kappa| \ll |\lambda|, \bar{g}$ and we consider as before two regimes 
for $s$. However the discussion will be much more involved. We give hereafter
the main steps.

\begin{itemize}
\item[(a)]{ $s \lsim m, A$:}
%{\bf (a) $s \lsim m, A$:}

Since  $|\kappa| s \ll m, A$ in this case, it is reliable to expand in the 
small parameter
$\kappa s/M_{soft}$. Adding up Eq.(\ref{ewsb1}) divided by $h_1$  to
Eq.(\ref{ewsb2}) divided by $h_2$ and using 
Eqs.(\ref{ewsb3}, \ref{Zmass}) to eliminate the combinations
$h_1 h_2$ and $h_1^2 + h_2^2$ on finds

\begin{equation}
2 \hat{\mu}^2 - x \hat{\mu} -\eta  + 
O( ( \frac{\kappa s}{M_{soft}})^2 ) = 0  \label{mureg31}
\end{equation}

with
\begin{eqnarray}
x= \frac{\lambda \kappa}{\bar{g}^2} \frac{M_Z^2 A_{\lambda}^2}{m_s^2 + \frac{\lambda^2}{\bar{g}^2} M_Z^2}( \frac{3}{A_\lambda} - \frac{A_\kappa}{m_s^2 + \frac{\lambda^2}{\bar{g}^2} M_Z^2}) \nonumber \\
\eta=  \frac{\lambda^2}{\bar{g}^2}(\frac{A_{\lambda}^2}{m_s^2 + \frac{\lambda^2}{\bar{g}^2}M_Z^2} - 1 ) M_Z^2 - m_1^2 - m_2^2 \label{eqeta}
\end{eqnarray}

Thus 
\begin{equation}
\hat{\mu} = \pm \sqrt{\frac{\eta}{2}} + O(x) \label{muconsist1}
\end{equation}
provided $\eta \gsim 0$. 

On the other hand, Eq.(\ref{kslam}) yields

\begin{equation}
(\frac{1}{t_\beta} + \frac{\kappa}{\lambda}) \hat{\mu}^2 +
A_{\lambda} \hat{\mu} + \frac{1}{2} ( \frac{ c_{2 \beta} M_Z^2 + 2 m_1^2}{t_\beta} + \frac{\lambda^2}{\bar{g}^2} s_{2 \beta} M_Z^2 ) =0 \label{mugeneral}
\end{equation}

Consistency between equations (\ref{mugeneral}) and (\ref{mureg31})
requires 

\begin{equation}
\hat{\mu} = - \frac{1}{2 t_\beta A_\lambda} (2 m_1^2 + \eta + (c_{2 \beta}
+ \frac{\lambda^2}{\bar{g}^2} ( 1 - c_{2 \beta}) ) M_Z^2 ) + O(x, 
\frac{\kappa}{\lambda}) \label{muconsist2}
\end{equation}

on top of Eq.(\ref{muconsist1}). Furthermore, combining 
Eqs.(\ref{eqh1}, \ref{eqh2}) to retrieve the familiar EWSB relation

\begin{equation}
\frac{M_Z^2}{2} = \frac{(m_1^2 +\hat{\mu}^2) - (m_2 + \hat{\mu}^2) t_\beta^2  
}{t_\beta^2 - 1} \label{ewsbmssm}
\end{equation}

and using Eq.(\ref{muconsist1}) one gets a further correlation between
$m_1, m_2, A_{\lambda}$ and $m_s$ in the form

\begin{equation}
m_1^2 - m_2^2 =  ((\frac{\lambda^2}{\bar{g}^2} - 1) M_Z^2 - X) c_{2 \beta} \label{correl0}
\end{equation}

where
\begin{equation}
X \equiv \frac{ \frac{\lambda^2}{\bar{g}^2} M_Z^2}{ m_s^2 +  \frac{\lambda^2}{\bar{g}^2} M_Z^2} A_\lambda^2 \label{eqX}
\end{equation}

Finally, equating the values $\hat{\mu}^2$ from 
Eqs.(\ref{muconsist1}, \ref{muconsist2}) 
and using Eq.(\ref{correl0}) to eliminate $m_2$ one finds the necessary
correlation among the mass parameters $m_1, m_s, A_{\lambda}$ and $M_Z$,
for given $t_\beta (\equiv \frac{h_2}{h_1}) $,

\begin{eqnarray}
\frac{X}{A_\lambda^2} &=& \frac{t_\beta^2}{1 - c_{2 \beta}} ( 1 \pm
\sqrt{ 1 - \frac{x_1}{A_\lambda^2}} ) \equiv {\cal R}^\pm \label{correl1} \\
A_\lambda^2 &\geq& x_1 \label{correl2} 
\end{eqnarray}

where

\begin{equation}
x_1 \equiv \frac{2}{t_\beta^2} ( 2 m_1^2 + ( \frac{\lambda^2}{\bar{g}^2} (1 - c_{2 \beta} ) + c_{2 \beta} )M_Z^2 ) \label{eqx1}
\end{equation}

Equation (\ref{correl1}) is easily obtained by eliminating $\hat{\mu}$
between Eqs.(\ref{muconsist1}) and (\ref{muconsist2}) and solving for
the resulting quadratic equation in $X$.  
(For simplicity, we skip from now on the explicit indication that all the 
relations are valid only to zero$^{th}$ order in small parameters like 
$\kappa/\lambda, \kappa/\bar{g}, \kappa s/M_{soft}$.)

Incidentally, we note here that $\eta \gsim 0$ as required by Eq.(\ref{muconsist1}), is automatically
implied by the correlations Eqs.(\ref{correl0}, \ref{correl1}). 
Furthermore, $m_s^2>0$ implies immediately

\begin{equation}
0 \le {\cal R}^\pm \le 1 \label{correl3}
\end{equation}

\noindent
from (\ref{eqX}) and (\ref{correl1}).
For definiteness we stick hereafter to the phenomenologically likely case 
$t_\beta >1 $. Then, it is easy to see from Eq.(\ref{correl1}) that
${\cal R}^+ <1 $ is excluded by $t_\beta >1 $, while ${\cal R}^-$ is acceptable 
and leads through Eq.(\ref{correl3}) to the constraint

\begin{equation}
A_\lambda^2 \geq \frac{x_1}{s_{2 \beta}^2} \label{correl4}
\end{equation}

which overpowers Eq.(\ref{correl2}).

One can now determine uniquely $\hat{\mu}$ from Eqs.(\ref{muconsist2}, \ref{eqeta}, \ref{correl0}, 
\ref{correl1}):

\begin{equation}
\hat{\mu} = -\frac{1-c_{2 \beta}}{2 t_{\beta}} \frac{X}{A_\lambda}=
-\frac{1}{2} t_{\beta} A_{\lambda} [ 1 - \sqrt{1 - \frac{x_1}{A_\lambda^2} } ]
\label{resmu}
\end{equation}

Now as far as $x_1$ ( Eq.(\ref{eqx1})) 
remains positive at the electroweak scale (which is the case if 
$\tan \beta$ does not become extremely large leading to substantially
negative $m_1^2$ at this same scale), equation (\ref{resmu}) shows that
$|\hat{\mu}|$ is a decreasing function of $|A_\lambda|$. Then using 
Eq.(\ref{correl4}) one gets the following upper bound on $\hat{\mu}^2$ 

\begin{equation}
\hat{\mu}^2 \le \frac{x_1}{4} \label{bound1}
\end{equation}

This constraint leads to an upper bound on the physical lightest
chargino mass, which may or may not be consistent with the
experimental lower bounds. To go further in this issue, let us include 
first the requirement that the electroweak symmetry breaking
extremum of the scalar potential should  indeed be  a minimum lower than the 
$SU(2)_L \times U(1)_Y$ symmetric one at vanishing scalar fields, i.e. that  

\begin{equation}
V_{min} < 0 \label{stabili}
\end{equation}

Evaluating Eq.(\ref{potmin}) at $\kappa \simeq 0$ and $\hat{\mu}$ as given by
 Eq.(\ref{resmu}), the above condition leads to

\begin{equation}
4 t_\beta^2 ( 1 + t_\beta^2) (x_2 - ( 1 + t_\beta^2) x_1 ) A_\lambda^2 +
(x_2 -  ( 1 + t_\beta^2)^2 x_1 )^2 > 0 \label{stabi}
\end{equation}

where $x_1$ is as given before, (Eq.(\ref{eqx1}),
 
\begin{equation}
x_2 = 4 (\frac{\lambda^2}{\bar{g}^2} + \frac{ c_{2 \beta}^2}{s_{2 \beta}^2}) M_Z^2 \label{eqx2}
\end{equation}

and Eq.(\ref{correl4}) has been implicitly used. 
As can be seen from Eq.(\ref{stabi}),
this equation leads to a constraint only if $x_2 - ( 1 + t_\beta^2) x_1 < 0$.
We need to consider two cases:\\

{\sl i)} $x_2 - ( 1 + t_\beta^2) x_1 > 0$: in this case Eq.(\ref{stabili})
is trivially verified, but Eq.(\ref{bound1}) together with 
$x_2 - ( 1 + t_\beta^2) x_1 > 0$ and (\ref{eqx2}) lead to

\begin{equation}
\hat{\mu}^2 \le \frac{ 1 + t_\beta^2}{4 t_\beta^2} (\frac{\lambda^2}{\bar{g}^2}
s_{2 \beta}^2 + c_{2 \beta}^2 ) M_Z^2 \label{bound2}
\end{equation} 

\noindent
Since the chargino sector is identical to that of the MSSM one can study
directly the effect of the above bound on the mass  of the 
lightest chargino denoted hereafter by $M_{\chi_1^+}$. 
Using a rigorous upper bound on $M_{\chi_1^+}$

\begin{equation}
M_{\chi_1^+} \le \sqrt{ \hat{\mu}^2 + \frac{ 2 M_W^2}{1 +t_\beta^2} }
\label{supbound}
\end{equation}

\noindent
in conjunction with Eq.(\ref{bound2}) one gets immediately a $\tan \beta$
dependent (and $\hat{\mu}$ independent) upper bound
on $M_{\chi_1^+}$. Confronting this upper bound with the present 
experimental lower bounds from the LEPII exclusion analyses \cite{LEP, LEP1}
one finds typically that for $\tan \beta >1$ only a small range of
$\tan \beta$ values close to $1$ are possible.    
For example, taking $\lambda = \bar{g} $ one has

\begin{equation}
M_{\chi_1^+} \le \sqrt{\frac{ 2}{1 +t_\beta^2} M_W^2  + 
\frac{ 1 + t_\beta^2}{4 t_\beta^2}  M_Z^2} \label{bound2p}
\end{equation}

\noindent
Comparing for instance with the experimental analysis 
of \cite{LEP1} (Fig 5 therein) 
one finds that only a very small range, $1< \tan \beta< 1.3-1.6$, is 
consistent with our regime in this case. Of course one should keep in mind
the model dependence of the experimental analyses. However we should stress
that since in the regime under consideration $\lambda$ is 
relatively large ($ \sim \bar{g}$), we are in a configuration which is safe
from significant unconventional signals due to displaced vertices that can 
occur in the (M+1)SSM \cite{EH2}. The comparison with MSSM-based experimental 
analyses is thus fully consistent here.\\

\noindent
{\sl ii)} $x_2 - ( 1 + t_\beta^2) x_1 < 0$: In this case the stability
of the electroweak symmetry breaking minimum requires 
that $A_\lambda^2$ be bounded from above at the electroweak scale.
This upper bound $\bar{A_\lambda^2}$ can be easily read from 
Eq.(\ref{stabi})\footnote{ which in turn leads to a phenomenologically harmless 
lower bound on $|\hat{\mu}|$, $|\hat{\mu}| >|\hat{\mu}(\bar{A_\lambda})|$, 
rather than an improved upper bound like in {\sl i)}.}. 
However, $A_\lambda^2<\bar{A_\lambda^2}$ will lead, through Eqs.(\ref{correl1},
\ref{eqX}, \ref{eqx1}, \ref{eqx2}) to an upper bound on $m_s^2$
(remember that Eq.(\ref{correl1}) is valid for ${\cal R}^-$, see
the discussion following Eq.(\ref{correl3})). 
Working out the algebra one finds

\begin{equation}
m_s^2< \frac{\lambda^2}{\bar{g}^2} \frac{M_Z^4}{4 m_1^2 - (t_\beta^2 - 1)M_Z^2}
\frac{t_\beta^2}{ ( 1 + t_\beta^2)^2} ((t_\beta^2 -1)^2 + 
4 t_\beta^2 \frac{\lambda^2}{\bar{g}^2} ) \label{correl5}
\end{equation}

\noindent
This inequality shows an anti-correlation between $m_s^2$ and $m_1^2$
which suggests that universality assumptions between the singlet and doublet 
soft scalar masses may be disfavoured. To see this for any value of $t_\beta$
one should plug the running expressions for $m_1^2, m_s^2$ given
in Eqs.(\eq{m1soft}, \eq{mSsoft}) and evaluate the auxiliary functions given
in Eqs.(\eq{ut} -- \eq{ukappa}) and in Appendix B (approximating them for
instance to their first order iteration). Instead, let us give here
a simpler and more qualitative argument. It is clear from Eqs.(\ref{bound1},
\ref{bound2}) and from the dependence of $x_1$ on $t_\beta$ (Eq.(\ref{eqx1})) 
that the higher the experimental exclusion bound on $M_{\chi_1^+}$ the less 
favoured are the large $t_\beta$ configurations. 
This conclusion is reinforced by the 
fact that for a given initial condition $m_0^2$, the running $m_1^2$ decreases 
faster for larger $t_\beta$, thus driving $x_1$ to small values. 
So let us concentrate only on the small $t_\beta$ region.  In this case
the running  $m_s^2$ and $m_1^2$ are well approximated by the simple analytical
solutions given in \cite{ERS2}, namely $m_1^2 \simeq m_0^2 + M_0^2/2,  
m_s^2 \simeq m_0^2$, in the vicinity of the infrared effective fixed point
for small $\tan \beta$ ($\sim 1$) and assuming universal initial conditions
for the scalar  and fermion soft masses (denoted respectively by
$m_0$  and $M_0$). Using these relations in Eq.(\ref{correl5}) one finds

\begin{equation}
m_0^2 < \frac{1}{2} ( -\frac{M_0^2}{2} + \sqrt{\frac{M_0^4}{4} + M_Z^4})
   + O( t_\beta -1) <  \frac{M_Z^2}{2} + O( t_\beta -1)
\end{equation}

\noindent
that is $m_0^2$ is forced to be rather small (eventually vanishing) 
for large $M_0^2$, the latter
being required by experimental lower bounds on  $M_{\chi_1^+}$. A way to
avoid such an anti-correlation is to relax the universality between
the singlet and doublet Higgs soft masses. 
      
\item[(b)]{ $s \gg m, A$:}
%{\bf (b) $s \gg m, A$:}

In this case, no generic statement can be made about the size of $\kappa s$.
If the magnitude of $\lambda$  is such that $M_{soft}/(\lambda s)$ 
is small then Eqs.(\ref{eqh1}, \ref{eqh2}) can be cast in the form

\begin{eqnarray}
h_1^2 &=& \frac{M_Z^2}{2 \bar{g}^2} [ 1 + \frac{ m_2^2- m_1^2}{2 \hat{\mu}^2}
+ O((\frac{M_{soft}}{\hat{\mu}})^4)] \\
h_2^2 &=&  \frac{M_Z^2}{2 \bar{g}^2} [ 1 + \frac{ m_1^2- m_2^2}{2 \hat{\mu}^2})
+ O((\frac{M_{soft}}{\hat{\mu}})^4) ] 
\end{eqnarray}

Feeding the above equations back in Eq.(\ref{ewsb1}) one gets,

\begin{equation}
\hat{\mu}^2 + \frac{\gamma}{\hat{\mu}} + \delta  + O(\frac{\kappa}{\lambda} (\frac{M_{soft}}{\hat{\mu}})^2, (\frac{M_{soft}}{\hat{\mu}})^4) \times \frac{M_Z^2}{2 \bar{g}^2} = 0
\label{mureg32}
\end{equation}

where

\begin{eqnarray}
\gamma &=& \frac{1}{2} A_{\lambda} ( m_1^2 - m_2^2) \\
\delta  &=& \frac{\lambda^2}{\bar{g}^2} M_Z^2 + m_1^2 + \frac{\kappa}{2 \lambda} (m_1^2 - m_2^2) 
\end{eqnarray}

In the deep regime 3, $|\kappa/\lambda| \ll 1$, 
thus $\delta$ is positive (except for very large $t_\beta$ where
$m_1^2$ can become negative at the electroweak scale).  It then follows from 
Eq.(\ref{mureg32}) that this regime can not be dynamically consistent in the 
case at hand, {\sl i.e.} as far as $\lambda$ is not too small in this regime
 so that $s \gg m, A$ implies very large $|\lambda| s (= \hat{\mu})$. 

Finally let us stress that Eq.(\ref{mureg32}) is more general than in the
context of regime 3 (the condition $\kappa \ll \bar{g}$ was not used in
establishing this equation). Furthermore,  
$|\delta|$ is by definition of order $M_{soft}^2$ whatever its sign.
Thus if $s$ happens to be extremely large compared to the soft masses, then
Eq.(\ref{mureg32}) forbids $\hat{\mu}$ to be very large too,
{\sl i.e.} $\lambda$ cannot be large ($\sim \bar{g}$ say). We thus
have a further ingredient in understanding purely numerical studies were
indeed very large $s$ required very small $\lambda$
(see for instance table 1 of reference \cite{ERS2}). 
\end{itemize}

\subsection{Summary for regime 3}
Regime 3 is the trickiest. We have shown that an $s$ much lager than
the soft parameters is forbidden unless the hierarchy 
$|\kappa| \ll |\lambda| \lsim M_{soft}/s$ is realized. On the other hand,
in the case where $s$ is of the order of the soft parameters
the stability of the EWSB vacuum has to be invoked. 
In configurations where this stability is automatic, $\hat{\mu}$ leads
to light charginos inconsistent with the present experimental limits
(except for a small window $1< \tan \beta <1.3 -1.6$ which will also be closed
by a few GeV improvement of these limits). When the stability of the EWSB
vacuum is not automatic, then the generic price to pay is small $\tan \beta$
values and either a relaxation of universality between the singlet and doublet 
Higgs soft scalar masses, or a small universal soft scalar mass $m_0$
anti-correlated with a large universal soft gaugino mass $M_0$ to
ensure consistency with experimental limits on the lightest chargino.
Clearly future improvement of these mass limits will reinforce the
above conclusions.

\section{Conclusion}

We have made an extended analytical study of the scale evolution
of the various basic parameters, as well as of the spontaneous electroweak 
symmetry breaking and the dynamical determination of the $\hat{\mu}$ parameter
in the (M+1)SSM. In particular, we identified four regimes of effective
infrared fixed points behaviour corresponding to various relative magnitudes
of the two couplings $\lambda$ and $\kappa$ that enter the gauge singlet scalar
sector.  These regimes correspond to the configurations of least sensitivity
to the initial (GUT-scale) conditions of most of the parameters.
We have determined analytically this sensitivity and shown how it
generalizes the MSSM case. Furthermore, the analysis of electroweak symmetry
breaking (which did not require any numerical scan over the parameter
space) showed that some of these regimes, taken in a wider sense, are
generically excluded by negative experimental searches for charginos 
or by purely dynamical considerations, and that the others lead to very large 
singlet vacuum expectation values which can be reduced, however, if some 
amount of non-universality of the soft parameters is allowed.\\ 

The general analytical solutions given in this paper do not rely on any 
model-dependent GUT-scale assumptions. 
In practice they lead to approximate solutions to the RGE's
(with controllable convergence), 
in analytically closed forms and for any value of 
the $\tan \beta$ parameter. They thus allow a precise study of the dynamical 
properties of the (M+1)SSM. Besides, they are readily generalizable to 
further extensions of the Higgs sector. This provides a basis to 
study the genericity of these properties beyond the (M+1)SSM and to gain a more
thorough understanding of the sensitivity of the Higgs sector phenomenology
to specific underlying supersymmetric models. \\

\noindent
{\sl Acknowledgments} \\
This work was done in the context of {\sl GDR-Supersym\'etrie}.
We would like to thank Ulrich Ellwanger and Cyril Hugonie for
useful discussions. Y.M. acknowledges financial support from MNESR.

%\newpage
\section*{Appendix A: The coefficients $a_{ki}$, $b_i$  and $c_{ki}$}
\renewcommand{\theequation}{A.\arabic{equation}}
\setcounter{equation}{0}

In this appendix we 
define the coefficients $a_{ki}$ and $c_{ki}$ introduced
in the RGE \eq{rge}

$$\vbox{\offinterlineskip \halign{
\tv# & \cc{#} & \tv# & \cc{#}  & \tv# &
\cc{#} & \tv# & \cc{#} & \tv# & \cc{#}& \tv# & \cc{#}& \tv# \cr
\noalign{\hrule}
&\cc{$a_{t t}=6$}&&\cc{$a_{bt}=1$} &&\cc{$a_{\tau t}=0$}&&\cc{$a_{\lambda t}=3$}
&&\cc{$a_{\kappa t}=0$ }& \cr
\noalign{\hrule}
&\cc{$a_{t b}=1$}&&\cc{$a_{bb}=6$} &&\cc{$a_{\tau b}=3$}&&\cc{$a_{\lambda b}=3$}
&&\cc{$a_{\kappa b}=0$ }& \cr
\noalign{\hrule}
&\cc{$a_{t \tau}=0$}&&\cc{$a_{b \tau}=1$} &&\cc{$a_{\tau \tau}=4$}&&\cc{$a_{\lambda \tau}=1$}
&&\cc{$a_{\kappa b}=0$ }& \cr
\noalign{\hrule}
&\cc{$a_{t \lambda}=1$}&&\cc{$a_{b \lambda}=1$} &&\cc{$a_{\tau \lambda}=1$}&&\cc{$a_{\lambda \lambda}=4$}
&&\cc{$a_{\kappa \lambda}=6$ }& \cr
\noalign{\hrule}
&\cc{$a_{t \kappa}=0$}&&\cc{$a_{b \kappa}=0$} &&\cc{$a_{\tau \kappa}=0$}&&\cc{$a_{\lambda \kappa}=2$}
&&\cc{$a_{\kappa \kappa}=6$ }& \cr
\noalign{\hrule}
}}$$

$$\vbox{\offinterlineskip \halign{
\tv# & \cc{#} & \tv# & \cc{#}  & \tv# &
\cc{#} & \tv# & \cc{#} & \tv# & \cc{#}& \tv# & \cc{#}& \tv# \cr
\noalign{\hrule}
&\cc{$b_{ 1}=11$}&&\cc{$b_{2}=1$} &&\cc{$b_{3}=-3$}& \cr
\noalign{\hrule}
}}$$

$$\vbox{\offinterlineskip \halign{
\tv# & \cc{#} & \tv# & \cc{#}  & \tv# &
\cc{#} & \tv# & \cc{#} & \tv# & \cc{#}& \tv# & \cc{#}& \tv# \cr
\noalign{\hrule}
&\cc{$c_{t 1}=\frac{13}{9}$}&&\cc{$c_{b1}=\frac{7}{9}$} &&\cc{$c_{\tau 1}=3$}&&\cc{$c_{\lambda 1}=1$}
&&\cc{$c_{\kappa 1}=0$ }& \cr
\noalign{\hrule}
&\cc{$c_{t 2}=3$}&&\cc{$c_{b2}=3$} &&\cc{$c_{\tau 2}=3$}&&\cc{$c_{\lambda 2}=3$}
&&\cc{$c_{\kappa 2}=0$ }& \cr
\noalign{\hrule}
&\cc{$c_{t 3}=\frac{16}{3}$}&&\cc{$c_{b 3}=\frac{16}{3}$} &&\cc{$c_{\tau 3}=0$}&&\cc{$c_{\lambda 3}=0$}
&&\cc{$c_{\kappa 3}=0$ }& \cr
\noalign{\hrule}
}}$$

\section*{Appendix B: Exact solutions for the $A$'s 
and the $\Sigma's$}
\renewcommand{\theequation}{B.\arabic{equation}}
\setcounter{equation}{0}
Hereafter we give the (recursive)-equations defining
the auxiliary functions which enter the $A$'s and $\Sigma$'s.

\vskip .5 truecm
\subsection*{The $A$'s}
\beq
e_t=\frac{1}{E_t}\frac{d\tilde E_t}{d\eta}
+Y_b^0\frac{A_b^0 \int u_b -\int u_be_b}{1+6Y^0_b\int u_b}
+Y_{\lambda}^0\frac{A_{\lambda}^0 \int u_{\lambda} -\int u_{\lambda}e_{\lambda}}{1+4Y^0_{\lambda}\int u_{\lambda}}
\label{eq:et}
\eeq

\bea
e_b&=&\frac{1}{E_b}\frac{d\tilde E_b}{d\eta}
+Y_t^0\frac{A_t^0 \int u_t -\int u_te_t}{1+6Y^0_t\int u_t}
+Y_{\tau}^0\frac{A_{\tau}^0 \int u_{\tau} -\int u_{\tau}e_{\tau}}
{1+4Y^0_{\tau}\int u_{\tau}}
\nonumber
\\
&+&Y_{\lambda}^0\frac{A_{\lambda}^0 
\int u_{\lambda} -\int u_{\lambda}e_{\lambda}}
{1+4Y^0_{\lambda}\int u_{\lambda}}
\label{eq:eb}
\eea

\beq
e_{\tau}=\frac{1}{E_{\tau}}\frac{d\tilde E_{\tau}}{d\eta}
+3Y_b^0\frac{A_b^0 \int u_b -\int u_be_b}{1+6Y^0_b\int u_b}
+Y_{\lambda}^0\frac{A_{\lambda}^0 
\int u_{\lambda} -\int u_{\lambda}e_{\lambda}}
{1+4Y^0_{\lambda}\int u_{\lambda}}
\eeq

\bea
e_{\lambda}&=&\frac{1}{E_{\lambda}}\frac{d\tilde E_{\lambda}}{d\eta}
+3Y_t^0\frac{A_t^0 \int u_t -\int u_te_t}{1+6Y^0_t\int u_t}
+3Y_b^0\frac{A_b^0 \int u_b -\int u_be_b}{1+6Y^0_b\int u_b}
\nonumber
\\
&+&Y_{\tau}^0\frac{A_{\tau}^0 \int u_{\tau} -\int u_{\tau}e_{\tau}}
{1+4Y^0_{\tau}\int u_{\tau}}
+2Y_{\kappa}^0\frac{A_{\kappa}^0 
\int u_{\kappa} -\int u_{\kappa}e_{\kappa}}
{1+6Y^0_{\kappa}\int u_{\kappa}}
\eea

\beq
e_{\kappa}=\frac{1}{E_{\kappa}}\frac{d\tilde E_{\kappa}}{d\eta}
+6Y_{\lambda}^0\frac{A_{\lambda}^0 
\int u_{\lambda} -\int u_{\lambda}e_{\lambda}}
{1+4Y^0_{\lambda}\int u_{\lambda}}
\label{eq:ekappa}
\eeq

\noindent
where the variations of $\tilde{E}_k$ should be taken at $\eta = \bar
\eta=0$ and are given by
\beq
\frac{1}{E_k}\frac{d\tilde E_k}{d \eta}=t\sum \limits_{i=1}^3c_{ki}
\alpha_iM_i^0 \label{eq:dtildeE}
\eeq

\vskip.5truecm
\subsection*{The $\Sigma$'s}

\bea
\xi_t&=&\frac{1}{E_t}\frac{d^2\tilde E_t}{d\eta d\overline{\eta}}
+\frac{2}{E_t}\frac{d\tilde E_t}{d\eta}
\left[
Y_b^0\frac{A_b^0 \int u_b -\int u_be_b}{1+6Y^0_b\int u_b}
+Y_{\lambda}^0\frac{A_{\lambda}^0 
\int u_{\lambda} -\int u_{\lambda}e_{\lambda}}
{1+4Y^0_{\lambda}\int u_{\lambda}}\right]
\nonumber
\\
&-&Y^0_b\frac{(\Sigma^0_b+(A^0_b)^2)\int u_b -2A^0_b \int u_be_b +\int u_b \xi_b}
{1+6Y^0_b\int u_b}
\nonumber
\\
&-&Y^0_{\lambda}\frac{(\Sigma^0_{\lambda}+(A^0_{\lambda})^2)\int u_{\lambda} -2A^0_{\lambda}
 \int u_{\lambda}e_{\lambda} +\int u_{\lambda} \xi_{\lambda}}
{1+4Y^0_{\lambda}\int u_{\lambda}}
\nonumber
\\
&+&7(Y_b^0)^2\left[\frac{A_b^0 \int u_b -\int u_be_b}{1+6Y^0_b\int u_b}\right]^2
+5(Y_{\lambda}^0)^2\left[\frac{A_{\lambda}^0 \int u_{\lambda} -\int u_{\lambda}
e_{\lambda}}{1+4Y^0_{\lambda}\int u_{\lambda}}\right]^2
\nonumber
\\
&+&2Y_b^0Y^0_{\lambda}\frac{(A_b^0 \int u_b -\int u_be_b)}{(1+6Y^0_b\int u_b)}
\frac{(A_{\lambda}^0 
\int u_{\lambda} -\int u_{\lambda}e_{\lambda})}
{(1+4Y^0_{\lambda}\int u_{\lambda})}
\label{eq:xit}
\eea

\bea
\xi_b&=&\frac{1}{E_b}\frac{d^2\tilde E_b}{d\eta d\overline{\eta}}
+\frac{2}{E_b}\frac{d\tilde E_b}{d\eta}
\left[
Y_t^0\frac{A_t^0 \int u_t -\int u_te_t}{1+6Y^0_t\int u_t}
+Y_{\tau}^0\frac{A_{\tau}^0 
\int u_{\tau} -\int u_{\tau}e_{\tau}}
{1+4Y^0_{\tau}\int u_{\tau}} \right.
\nonumber
\\
&+&\left.Y_{\lambda}^0\frac{A_{\lambda}^0 
\int u_{\lambda} -\int u_{\lambda}e_{\lambda}}
{1+4Y^0_{\lambda}\int u_{\lambda}}\right]
\nonumber
\\
&-&Y^0_t\frac{(\Sigma^0_t+(A^0_t)^2)\int u_t -2A^0_t \int u_te_t +\int u_t \xi_t}
{1+6Y^0_t\int u_t}
\nonumber
\\
&-&Y^0_{\tau}\frac{(\Sigma^0_{\tau}+(A^0_{\tau})^2)\int u_{\tau} -2A^0_{\tau}
 \int u_{\tau}e_{\tau} +\int u_{\tau} \xi_{\tau}}
{1+4Y^0_{\tau}\int u_{\tau}}
\nonumber
\\
&-&Y^0_{\lambda}\frac{(\Sigma^0_{\lambda}+(A^0_{\lambda})^2)\int u_{\lambda} -2A^0_{\lambda}
 \int u_{\lambda}e_{\lambda} +\int u_{\lambda} \xi_{\lambda}}
{1+4Y^0_{\lambda}\int u_{\lambda}}
\nonumber
\\
&+&7(Y_t^0)^2\left[\frac{A_t^0 \int u_t -\int u_te_t}{1+6Y^0_t\int u_t}\right]^2
\nonumber
\\
&+&5(Y_{\tau}^0)^2\left[\frac{A_{\tau}^0 \int u_{\tau} -\int u_{\tau}
e_{\tau}}{1+4Y^0_{\tau}\int u_{\tau}}\right]^2
+5(Y_{\lambda}^0)^2\left[\frac{A_{\lambda}^0 \int u_{\lambda} -\int u_{\lambda}
e_{\lambda}}{1+4Y^0_{\lambda}\int u_{\lambda}}\right]^2
\nonumber
\\
&+&2Y_t^0Y^0_{\lambda}\frac{(A_t^0 \int u_t -\int u_te_t)}{(1+6Y^0_t\int u_t)}
\frac{(A_{\lambda}^0 
\int u_{\lambda} -\int u_{\lambda}e_{\lambda})}
{(1+4Y^0_{\lambda}\int u_{\lambda})}
\nonumber
\\
&+&2Y_t^0Y^0_{\tau}\frac{(A_t^0 \int u_t -\int u_te_t)}{(1+6Y^0_t\int u_t)}
\frac{(A_{\tau}^0 
\int u_{\tau} -\int u_{\tau}e_{\tau})}
{(1+4Y^0_{\tau}\int u_{\tau})}
\nonumber
\\
&+&2Y_{\tau}^0Y^0_{\lambda}\frac{(A_{\tau}^0 \int u_{\tau} -
\int u_{\tau}e_{\tau}
)}{(1+4Y^0_{\tau}\int u_{\tau})}
\frac{(A_{\lambda}^0 
\int u_{\lambda} -\int u_{\lambda}e_{\lambda})}
{(1+4Y^0_{\lambda}\int u_{\lambda})}
\label{eq:xib}
\eea

\bea
\xi_{\tau}&=&\frac{1}{E_{\tau}}\frac{d^2\tilde E_{\tau}}
{d\eta d\overline{\eta}}
+\frac{2}{E_{\tau}}\frac{d\tilde E_{\tau}}{d\eta}
\left[
3Y_{b}^0\frac{A_{b}^0 
\int u_{b} -\int u_{b}e_{b}}
{1+6Y^0_{b}\int u_{b}}
+Y_{\lambda}^0\frac{A_{\lambda}^0 
\int u_{\lambda} -\int u_{\lambda}e_{\lambda}}
{1+4Y^0_{\lambda}\int u_{\lambda}}\right]
\nonumber
\\
&-&3\frac{(\Sigma^0_b+(A^0_b)^2)\int u_b -2A^0_b \int u_be_b +\int u_b \xi_b}
{1+6Y^0_b\int u_b}
\nonumber
\\
&-&Y^0_{\lambda}\frac{(\Sigma^0_{\lambda}+(A^0_{\lambda})^2)\int u_{\lambda} -2A^0_{\lambda}
 \int u_{\lambda}e_{\lambda} +\int u_{\lambda} \xi_{\lambda}}
{1+4Y^0_{\lambda}\int u_{\lambda}}
\nonumber
\\
&+&27(Y_b^0)^2\left[\frac{A_b^0 \int u_b -\int u_be_b}
{1+6Y^0_b\int u_b}\right]^2
+5(Y_{\lambda}^0)^2\left[\frac{A_{\lambda}^0 \int u_{\lambda} -\int u_{\lambda}
e_{\lambda}}{1+4Y^0_{\lambda}\int u_{\lambda}}\right]^2
\nonumber
\\
&+&6Y_b^0Y^0_{\lambda}\frac{(A_b^0 \int u_b -\int u_be_b)}
{(1+6Y^0_b\int u_b)}
\frac{(A_{\lambda}^0 
\int u_{\lambda} -\int u_{\lambda}e_{\lambda})}
{(1+4Y^0_{\lambda}\int u_{\lambda})},
\eea

\bea
\xi_{\lambda}&=&\frac{1}{E_{\lambda}}\frac{d^2\tilde E_{\lambda}}{d\eta d\overline{\eta}}
+\frac{2}{E_{\lambda}}\frac{d\tilde E_{\lambda}}{d\eta}
\left[
3Y_t^0\frac{A_t^0 \int u_t -\int u_te_t}{1+6Y^0_t\int u_t}
+3Y_b^0\frac{A_b^0 \int u_b -\int u_be_b}{1+6Y^0_b\int u_b}\right.
\nonumber
\\
&+&\left.Y_{\tau}^0\frac{A_{\tau}^0 
\int u_{\tau} -\int u_{\tau}e_{\tau}}
{1+4Y^0_{\tau}\int u_{\tau}}
+2Y_{\kappa}^0\frac{A_{\kappa}^0 
\int u_{\kappa} -\int u_{\kappa}e_{\kappa}}
{1+6Y^0_{\kappa}\int u_{\kappa}}
\right]
\nonumber
\\
&-&3Y^0_t\frac{(\Sigma^0_t+(A^0_t)^2)\int u_t -2A^0_t \int u_te_t +\int u_t \xi_t}
{1+6Y^0_t\int u_t}
\nonumber
\\
&-&3Y^0_b\frac{(\Sigma^0_b+(A^0_b)^2)\int u_b -2A^0_b \int u_be_b +\int u_b \xi_b}
{1+6Y^0_b\int u_b}
\nonumber
\\
&-&Y^0_{\tau}\frac{(\Sigma^0_{\tau}+(A^0_{\tau})^2)\int u_{\tau} -2A^0_{\tau}
 \int u_{\tau}e_{\tau} +\int u_{\tau} \xi_{\tau}}
{1+4Y^0_{\tau}\int u_{\tau}}
\nonumber
\\
&-&2Y^0_{\kappa}\frac{(\Sigma^0_{\kappa}+(A^0_{\kappa})^2)\int u_{\kappa} -2A^0_{\kappa}
 \int u_{\kappa}e_{\kappa} +\int u_{\kappa} \xi_{\kappa}}
{1+6Y^0_{\kappa}\int u_{\kappa}}
\nonumber
\\
&+&27(Y_t^0)^2\left[\frac{A_t^0 \int u_t -\int u_te_t}{1+6Y^0_t\int u_t}\right]^2
+27(Y_b^0)^2\left[\frac{A_b^0 \int u_b -\int u_be_b}{1+6Y^0_b\int u_b}\right]^2
\nonumber
\\
&+&5(Y_{\tau}^0)^2\left[\frac{A_{\tau}^0 \int u_{\tau} -\int u_{\tau}
e_{\tau}}{1+4Y^0_{\tau}\int u_{\tau}}\right]^2
+16(Y_{\kappa}^0)^2\left[\frac{A_{\kappa}^0 \int u_{\kappa} -\int u_{\kappa}
e_{\kappa}}{1+6Y^0_{\kappa}\int u_{\kappa}}\right]^2
\nonumber
\\
&+&18Y_t^0Y^0_{b}\frac{(A_t^0 \int u_t -\int u_te_t)}{(1+6Y^0_t\int u_t)}
\frac{(A_{b}^0 
\int u_{b} -\int u_{b}e_{b})}
{(1+6Y^0_{b}\int u_{b})}
\nonumber
\\
&+&6Y_t^0Y^0_{\tau}\frac{(A_t^0 \int u_t -\int u_te_t)}{(1+6Y^0_t\int u_t)}
\frac{(A_{\tau}^0 
\int u_{\tau} -\int u_{\tau}e_{\tau})}
{(1+4Y^0_{\tau}\int u_{\tau})}
\nonumber
\\
&+&12Y_t^0Y^0_{\kappa}\frac{(A_t^0 \int u_t -\int u_te_t)}{(1+6Y^0_t\int u_t)}
\frac{(A_{\kappa}^0 
\int u_{\kappa} -\int u_{\kappa}e_{\kappa})}
{(1+6Y^0_{\kappa}\int u_{\kappa})}
\nonumber
\\
&+&6Y_b^0Y^0_{\tau}\frac{(A_b^0 \int u_b -\int u_be_b)}{(1+6Y^0_b\int u_b)}
\frac{(A_{\tau}^0 
\int u_{\tau} -\int u_{\tau}e_{\tau})}
{(1+4Y^0_{\tau}\int u_{\tau})}
\nonumber
\\
&+&12Y_b^0Y^0_{\kappa}\frac{(A_b^0 \int u_b -\int u_be_b)}{(1+6Y^0_b\int u_b)}
\frac{(A_{\kappa}^0 
\int u_{\kappa} -\int u_{\kappa}e_{\kappa})}
{(1+6Y^0_{\kappa}\int u_{\kappa})}
\nonumber
\\
&+&4Y_{\tau}^0Y^0_{\kappa}\frac{(A_{\tau}^0 \int u_{\tau} -
\int u_{\tau}e_{\tau}
)}{(1+4Y^0_{\tau}\int u_{\tau})}
\frac{(A_{\kappa}^0 
\int u_{\kappa} -\int u_{\kappa}e_{\kappa})}
{(1+6Y^0_{\kappa}\int u_{\kappa})},
\eea

\bea
\xi_{\kappa}&=&\frac{1}{E_{\kappa}}\frac{d^2\tilde E_{\kappa}}
{d\eta d\overline{\eta}}
+\frac{2}{E_{\kappa}}\frac{d\tilde E_{\kappa}}{d\eta}
\left[
6Y_{\lambda}^0\frac{A_{\lambda}^0 
\int u_{\lambda} -\int u_{\lambda}e_{\lambda}}
{1+4Y^0_{\lambda}\int u_{\lambda}}\right]
\nonumber
\\
&-&6\frac{(\Sigma^0_{\lambda}+(A^0_{\lambda})^2)\int u_{\lambda} -2A^0_{\lambda}
 \int u_{\lambda}e_{\lambda} +\int u_{\lambda} \xi_{\lambda}}
{1+4^0_{\lambda}\int u_{\lambda}}
\nonumber
\\
&+&60(Y_{\lambda}^0)^2\left[\frac{A_{\lambda}^0 \int u_{\lambda} 
-\int u_{\lambda}
e_{\lambda}}{1+4Y^0_{\lambda}\int u_{\lambda}}\right]^2
\label{eq:xikappa}
\eea

\noindent 
with

\beq
\frac{1}{E_k}\frac{d^2\tilde E_k}{d \eta d\overline{\eta}}|_{\eta =0,
\bar \eta=0}=
t^2(\sum \limits_{i=1}^3c_{ki}\alpha_iM_i^0)^2
+2t\sum \limits_{i=1}^3c_{ki}\alpha_i(M_i^0)^2
-t^2\sum \limits_{i=1}^3c_{ki}\alpha_i^2(M_i^0)^2 \label{eq:d2tildeE}
\eeq

\vskip.5truecm
\subsection*{The $m$'s}
Due to linear relations 
\cite{Codoban, Moultaka} which follow from the RG equations
\eq{rge}, \eq{RGEm} (dropping out momentarily the trace term $S$
in the latter equation), one can express the soft masses 
 for squarks, sleptons and Higgses in terms of the $\Sigma_k$
in the form 
%\footnote{we have just written here the dependence on the ``matter'' 
%soft masses, not on the gaugino ones, its expression has no evident
%analytical solution. Furthermore, in these expressions we do not take
%into account the effect of the trace. This effect is given in 
%eq.[\ref{eq:mtrace}]}

\bea
m_T^2&=&(m_T^0)^2+\frac{27f_1-81f_2}{96} \label{eq:mTsoft}
\\
&+&\frac{45(\Sigma_t-\Sigma_t^0)+3(\Sigma_b-\Sigma_b^0)
+6(\Sigma_{\tau}-\Sigma_{\tau}^0)-27(\Sigma_{\lambda}-\Sigma_{\lambda}^0)
+9(\Sigma_{\kappa}-\Sigma_{\kappa}^0)}{96}\nonumber
\eea

\bea
m_B^2&=&(m_B^0)^2+\frac{21f_1-63f_2}{96}
\\
&+&\frac{3(\Sigma_t-\Sigma_t^0)+45(\Sigma_b-\Sigma_b^0)
+6(\Sigma_{\tau}-\Sigma_{\tau}^0)-21(\Sigma_{\lambda}-\Sigma_{\lambda}^0)
+7(\Sigma_{\kappa}-\Sigma_{\kappa}^0)}{96}
\nonumber
\eea

\bea
m_Q^2=(m_Q^0)^2&+&\frac{-24f_1+72f_2}{96}
\\
&+&\frac{24(\Sigma_t-\Sigma_t^0)+24(\Sigma_b-\Sigma_b^0)
-24(\Sigma_{\lambda}-\Sigma_{\lambda}^0)
+8(\Sigma_{\kappa}-\Sigma_{\kappa}^0)}{96}
\nonumber
\eea

\bea
m_L^2&=&(m_L^0)^2+\frac{-33f_1+99f_2}{96}
\\
&+&\frac{9(\Sigma_t-\Sigma_t^0)-9(\Sigma_b-\Sigma_b^0)
+30(\Sigma_{\tau}-\Sigma_{\tau}^0)-15(\Sigma_{\lambda}-\Sigma_{\lambda}^0)
+5(\Sigma_{\kappa}-\Sigma_{\kappa}^0)}{96}
\nonumber
\eea

\bea
m_E^2&=&(m_E^0)^2+\frac{30f_1-90f_2}{96}
\\
&+&\frac{18(\Sigma_t-\Sigma_t^0)-18(\Sigma_b-\Sigma_b^0)
+60(\Sigma_{\tau}-\Sigma_{\tau}^0)-30(\Sigma_{\lambda}-\Sigma_{\lambda}^0)
+10(\Sigma_{\kappa}-\Sigma_{\kappa}^0)}{96}
\nonumber
\eea

\bea
m_1^2&=&(m_1^0)^2+\frac{3f_1-9f_2}{96} \label{eq:m1soft}
\\
&+&\frac{-27(\Sigma_t-\Sigma_t^0)+27(\Sigma_b-\Sigma_b^0)
+6(\Sigma_{\tau}-\Sigma_{\tau}^0)+45(\Sigma_{\lambda}-\Sigma_{\lambda}^0)
-15(\Sigma_{\kappa}-\Sigma_{\kappa}^0)}{96}
\nonumber
\eea

\bea
m_2^2&=&(m_2^0)^2+\frac{-3f_1+9f_2}{96}
\\
&+&\frac{27(\Sigma_t-\Sigma_t^0)-27(\Sigma_b-\Sigma_b^0)
-6(\Sigma_{\tau}-\Sigma_{\tau}^0)+51(\Sigma_{\lambda}-\Sigma_{\lambda}^0)
-17(\Sigma_{\kappa}-\Sigma_{\kappa}^0)}{96}
\nonumber
\\
m_s^2&=&(m_s^0)^2+\frac{1}{3}(\Sigma_{\kappa}-\Sigma_{\kappa}^0)
\label{eq:mSsoft}
\eea

\noindent
where
$$f_i=\frac{(M_i^0)^2}{b_i}\left(1-\frac{1}{(1+b_i\alpha_0t)^2}\right).$$

\noindent
To obtain the complete solutions one should add to each of the
above equations the corresponding trace term contribution, see
Eqs.\eq{solm}, \eq{mtrace} and the discussion in section 3.

\section*{Appendix C: The $Y$'s in the IRQFP regimes }
\renewcommand{\theequation}{C.\arabic{equation}}
\setcounter{equation}{0}

We give in this appendix some details about the derivation of
Eqs.\eq{FPtbtau} --\eq{FPkappa}.  
%\subsection*{The Yukawa's}
For later use we define
\beq
r_i=\frac{Y^0_i}{Y^0_t} < \infty
\eeq
\noindent
when it makes sense, namely when $Y^0_t$ and $Y^0_i$ go 
simultaneously to infinity with a fixed ratio $r_i$.

\noindent
We will show inductively, in the same spirit as \cite{Moultaka}, that 
in such a regime  $u_i \to u_i^\infty$ with 

\beq
u_i^\infty=\frac{u^{\mathrm{FP}}_i}{(Y^0_t)^{p_i}}  \label{eq:uansatz}
\eeq

\noindent
where the $p_i$'s are fixed numbers which we will explicitly determine,
and the $u_i^{\mathrm{FP}}$'s are initial conditions independent and will be 
defined implicitly through equations
\eq{utFP} - \eq{ukappaFP}.

%determined  turn out to depend on the large
%Yukawa initial values only through ratios $r_i$
%In fact, as in \cite{Auberson}, we must first proved by a recurrence
%on the $p_i$'s that the solutions of the form (C.2) exist
%and is consistent with the following equations for the $u_k$'s.
%We then calculated their different values by finding the limit 
%of the following easy recurrent matrix system from the equation 
%\eq{ut}- \eq{ukappa}, where $n$ denote the order of the
%development. 

To proceed we consider finite order iteration approximations to 
Eqs.\eq{ut}- \eq{ukappa}. To obtain the $(n+1)^{th}$ order approximation
to $u_i$, denoted $u_i^{(n+1)}$, in terms of the $u_i^{(n)}$, one makes
the formal substitutions $u_i \to u_i^{(n+1)}$ and  $u_i \to u_i^{(n)}$
respectively on the lefthand and righthand sides of 
Eqs.\eq{ut}- \eq{ukappa}. If for any given one of the four IRQFP regimes
of section 3.2, the $u_i$'s have the  behaviour of Eq.\eq{uansatz} at the
$n^{th}$ order, i.e.

\beq
u_i^{(n)}=\frac{{u^{\mathrm{FP}}}^{(n)}_i}{(Y^0_t)^{p_i^{(n)}}} 
\eeq

\noindent
where the scale dependent functions ${u^{\mathrm{FP}}}^{(n)}$ are $Y^0_t$ 
independent but possibly $r_b,
r_\tau, r_\lambda, r_\kappa$ dependent, then the same is true at
the $(n+1)^{th}$ order with the following recursive equation
for the $p_i$'s

\beq
\left(
\begin{array}{l}
p_t^{(n+1)} \cr
p_b^{(n+1)} \cr
p_{\tau}^{(n+1)} \cr
p_{\lambda}^{(n+1)} \cr
p_{\kappa}^{(n+1)} \cr
\end{array}
\right)
=
\left(
\begin{array}{lllll}
0 & \frac{1}{6} & 0 &\frac{1}{4} & 0 \cr
\frac{1}{6} & 0 & \frac{1}{4} & \frac{1}{4} & 0 \cr
0 & \frac{1}{2} & 0 & \frac{1}{4} & 0 \cr
\frac{1}{2} & \frac{1}{2} & \frac{1}{4} & 0 & \frac{2}{3} \cr
0 & 0 & 0 & \frac{3}{2} & 0
\end{array}
\right)
\left(
\begin{array}{l}
(1 - p_t^{(n)}) \; \theta[1 - p_t^{(n)}] \cr
(1 - p_b^{(n)}) \; \theta[1 - p_b^{(n)}]\cr
(1 - p_{\tau}^{(n)}) \; \theta[1 - p_{\tau}^{(n)}] \cr
(1 - p_{\lambda}^{(n)}) \; \Delta_{34}[q, 1 - p_{\lambda}^{(n)}] \cr
(1 - p_{\kappa}^{(n)}) \; \Delta_{24}[q, 1 - p_{\kappa}^{(n)}]\cr
\end{array}
\right)
\label{eq:matrice}
\eeq

%\left(
%\begin{array}{lllll}
%0 & -\frac{1}{6} & 0 &-\frac{1}{4}\delta_{q3} & 0 \cr
%-\frac{1}{6} & 0 & -\frac{1}{4} & -\frac{1}{4}\delta_{q3} & 0 \cr
%0 & -\frac{1}{2} & 0 & -\frac{1}{4}\delta_{q3} & 0 \cr
%-\frac{1}{2} & -\frac{1}{2} & -\frac{1}{4} & 0 & 
%-\frac{2}{3}(\delta_{q2}+\delta_{q4}) \cr
%0 & 0 & 0 & -\frac{3}{2}\delta_{q3} & 0
%\end{array}
%\right)
%\left(
%\begin{array}{l}
%p_t^{(n)} \cr
%p_b^{(n)} \cr
%p_{\tau}^{(n)} \cr
%p_{\lambda}^{(n)} \cr
%p_{\kappa}^{(n)} \cr
%\end{array}
%\right)
%+
%\left(
%\begin{array}{l}
%\frac{1}{6}+\frac{1}{4}\delta_{q3} \cr
%\frac{5}{12} +\frac{1}{4} \delta_{q3} \cr
%\frac{1}{2} + \frac{1}{4} \delta_{q3} \cr
%\frac{5}{4} +\frac{2}{3}(\delta_{q2}+\delta_{q4}) \cr
%\frac{3}{2}\delta_{q3}
%\end{array}
%\right)
%\label{eq:matrice}
%\eeq

\noindent
where we define

\begin{equation}
\Delta_{ij}[q, x] \equiv (\delta_{qi} + \delta_{qj}) \; \theta[x]
\end{equation}
\noindent
the $\delta$'s are Kronecker's and $\theta$
is the Heaviside function.
Equation \eq{matrice} describes compactly the four IRQFP regimes
labeled by $q=1,2,3,4$. The $\theta$ function is here
to account in general for the fact  that some of the $p_i$'s can become
larger than one at some iteration order.  It is important to
keep this point under control since 
in a regime where some $Y^0_{i}$ is very large, 
one  has\footnote{provided that $t$ is also large enough.} 
$1+Y^0_{i}\int_0^t u_{i}^{(n)}\sim Y^0_{i}\int_0^t u_{i}^{(n)} $ if 
$p_{i}^{(n)}<1$ and $1+Y^0_{i}\int_0^t u_{i}^{(n)}\sim 1$ if $p_{i}^{(n)}>1$
 ( the critical value $p_{i}^{(n)}=1$ is never met).
Thus the proof of convergence to a unique form for the $u^\infty_i$'s
( {\sl i.e.} definite limiting values for the $p_i$'s and for the functions
$u^{\mathrm{FP}}_i$ ) needs some care. Direct numerical inspection of the
iterations in \eq{matrice} shows that
only $p_{\lambda}^{(n)}$ can be either bigger or smaller than one. 
For regimes 1,2,3, $p_{\lambda}^{(n)}$ is found to be constantly
smaller (or bigger) than one at each iteration. Regime 4 leads
to some slight complications: $p_{\lambda}^{(n)}$ and
$p_{\kappa}^{(n)}$ take alternating magnitudes, 
$p_{\lambda}^{(n)}>1, p_{\kappa}^{(n)}>0$
implying $p_{\lambda}^{(n+1)}<1, p_{\kappa}^{(n+1)}=0$ and vice versa
up to the first six iterations, but then stabilize with
$p_{\lambda}^{(n)}> 1$, $p_{\kappa}^{(n)}=0$ for all $n \ge 7$. Armed
with this numerical information, one can then consider Eq.\eq{matrice}
for any $n \ge 7$ to prove recursively that for all four regimes 
$0 < p_{t, b, \tau, \kappa}^{(n)} < 1$, while $0< p_{\lambda}^{(n)}<1$
for regimes 1,3 and $p_{\lambda}^{(n)}> 1$ for regimes 2, 4.
Eliminating the $\theta$ functions correspondingly and solving
Eq.\eq{matrice} for $p_i^{(n+1)}=p_i^{(n)}=p_i$, we find the limiting
values given in the Table 1.

\begin{table}
$$\vbox{\offinterlineskip \halign{
\tv# & \cc{#} & \tv# & \cc{#}  & \tv# &
\cc{#} & \tv# & \cc{#} & \tv# & \cc{#}& \tv# & \cc{#}& \tv# \cr
\noalign{\hrule}
& &&\cc{$Y^0_{\kappa}, Y^0_{\lambda} < \infty$ \bf{(1)}} &&\cc{$Y^0_{\lambda}< \infty$ \bf{(2)}}
&&\cc{$Y^0_{\kappa}< \infty$ \bf{(3)}}& &\cc{$/$ \bf{(4)}}&\cr
\noalign{\hrule}
&$p_t$&&\cc{$7/61$} &&\cc{$7/61$}&&\cc{$4/31$}
& &\cc{$7/61$}&\cr
\noalign{\hrule}
&\cc{$p_b$}&&\cc{$19/61$} &&\cc{$19/61$}&&\cc{$10/31$}
& &\cc{$19/61$}&\cr
\noalign{\hrule}
&\cc{$p_{\tau}$}&&\cc{$21/61$} &&\cc{$21/61$}&&\cc{$11/31$}
& &\cc{$21/61$}&\cr
\noalign{\hrule}
&\cc{$p_{\lambda}$}&&\cc{$58/61$} &&\cc{$296/183$}&&\cc{$29/31$}
& &\cc{$296/183$}&\cr
\noalign{\hrule}
&\cc{$p_{\kappa}$}&&\cc{$0$} &&\cc{$0$}&&\cc{$3/31$}
& &\cc{$0$}&\cr
\noalign{\hrule}
}}$$
\caption{Powers entering Eq. \eq{uansatz} in the various IRQFP regimes}
\end{table}

The same result can be derived through the more systematic method of solving 
directly the linear system \eq{matrice} for $p_i^{(n+1)}=p_i^{(n)}=p_i$, 
keeping formally the $\theta$ functions and not relying on any prior numerical 
information about the iterations.
Then, checking the consistency of the solutions for all possible combinations
of the $p_i$'s being greater or smaller than $1$, one delineates the
acceptable ones. It turns out that
there is only one consistent solution for each regime leading to the values
given in the table.

\noindent   
Knowing the $p_i$'s one can now extract the $u_i^{\mathrm{FP}}$'s,

\beq
u_t^{\mathrm{FP}}=\left\{
\begin{array}{ll}
\frac{E_t}{(6r_b)^{\frac{1}{6}}(\int u_b^{\mathrm{FP}})^{\frac{1}{6}}}
&
\mathrm{regimes~}1,2 \mathrm{~and~} 4 \cr
\frac{E_t}{(6r_b)^{\frac{1}{6}}(4r_{\lambda})^{\frac{1}{4}}(\int u_b^{\mathrm{FP}})^{\frac{1}{6}}(\int u_{\lambda}^{\mathrm{FP}})^{\frac{1}{4}}}
&
\mathrm{regime~}3 \cr
\end{array}  
\right.
\label{eq:utFP}
\eeq

\beq
u_b^{\mathrm{FP}}=\left\{
\begin{array}{ll}
\frac{E_b}{(6)^{\frac{1}{6}}(4r_{\tau})^{\frac{1}{4}}(\int u_t^{\mathrm{FP}})^{\frac{1}{6}}(\int u_{\tau}^{\mathrm{FP}})^{\frac{1}{4}}}
&
\mathrm{regimes~}1,2 \mathrm{~and~} 4 \cr
\frac{E_t}{(6)^{\frac{1}{6}}(4r_{\tau})^{\frac{1}{4}}(4r_{\lambda})^{\frac{1}{4}}(\int u_t^{\mathrm{FP}})^{\frac{1}{6}}
(\int u_{\tau}^{\mathrm{FP}})^{\frac{1}{4}}
(\int u_{\lambda}^{\mathrm{FP}})^{\frac{1}{4}}}
&
\mathrm{regime~} 3 \cr
\end{array} 
\right.
\eeq

\beq
u_{\tau}^{\mathrm{FP}}=\left\{
\begin{array}{ll}
\frac{E_{\tau}}{(6r_b)^{\frac{1}{2}}(\int u_b^{\mathrm{FP}})^{\frac{1}{2}}}
&
\mathrm{regimes~}1,2 \mathrm{~and~} 4 \cr
\frac{E_t}{(6r_b)^{\frac{1}{2}}(4r_{\lambda})^{\frac{1}{4}}(\int u_b^{\mathrm{FP}})^{\frac{1}{2}}(\int u_{\lambda}^{\mathrm{FP}})^{\frac{1}{4}}}
&
\mathrm{regime~} 3 \cr
\end{array} 
\right.
\eeq

\beq
u_{\lambda}^{\mathrm{FP}}=\left\{
\begin{array}{ll}
\frac{E_{\lambda}}{(6)^{\frac{1}{2}}
(6r_{b})^{\frac{1}{2}}
(4r_{\tau})^{\frac{1}{4}}
(\int u_t^{\mathrm{FP}})^{\frac{1}{2}}
(\int u_b^{\mathrm{FP}})^{\frac{1}{2}}
(\int u_{\tau}^{\mathrm{FP}})^{\frac{1}{4}}(1 + 6Y_\kappa^0 t)^{\frac{2}{3}}}
&
\mathrm{regime~}1  \cr
\frac{E_{\lambda}}{(6)^{\frac{1}{2}}
(6r_{b})^{\frac{1}{2}}
(4r_{\tau})^{\frac{1}{4}}
(\int u_t^{\mathrm{FP}})^{\frac{1}{2}}
(\int u_b^{\mathrm{FP}})^{\frac{1}{2}}
(\int u_{\tau}^{\mathrm{FP}})^{\frac{1}{4}}}
&
\mathrm{regime~}3 \cr
\frac{E_{\lambda}}{
(6)^{\frac{1}{2}}
(6r_{b})^{\frac{1}{2}}
(4r_{\tau})^{\frac{1}{4}}
(6r_{\kappa})^{\frac{2}{3}}
(\int u_t^{\mathrm{FP}})^{\frac{1}{2}}
(\int u_b^{\mathrm{FP}})^{\frac{1}{2}}
(\int u_{\tau}^{\mathrm{FP}})^{\frac{1}{4}}
(\int u_{\kappa}^{\mathrm{FP}})^{\frac{2}{3}}}
&
\mathrm{regimes~}2 \mathrm{~and~} 4 \cr
\end{array} 
\right.
\eeq

\beq
u_{\kappa}^{\mathrm{FP}}=\left\{
\begin{array}{ll}
1
&
\mathrm{regimes~}1, 2 \mathrm{~and~} 4 \cr \cr
\frac{E_{\kappa}}
{(4r_{\lambda})^{\frac{3}{2}}
(\int u_{\lambda}^{\mathrm{FP}})^{\frac{3}{2}}
}
&
\mathrm{regime~}3  \cr
\end{array} 
\right.
\label{eq:ukappaFP}
\eeq

Finally let us note that $p_{\kappa}$ is non vanishing (but small) only in 
regime 3.
Consequently, the evolution of $Y_{\kappa}$ will be very slow in this case, 
and a IRQFP regime will be theoretically obtained for very large values of 
$Y^0_{\kappa}$
(or very large values of $r_{\kappa}$), as it has been seen in the numerical
analysis of section 4.

\section*{Appendix D: The soft parameters in the IRQFP regimes}
\renewcommand{\theequation}{D.\arabic{equation}}
\setcounter{equation}{0} 

In this appendix we discuss in some detail the sensitivity of the soft
couplings and masses to their initial conditions in the IRQFP regimes.
Keeping in mind that one can derive the results directly from the
Yukawa couplings discussed in the previous appendix through the substitutions 
Eq.\eq{grassmann}, it is instructive to make a direct study starting off 
directly from the analytical forms in the soft sector.

\subsection*{The $A$'s}

To understand the behaviour of the soft trilinear couplings in the 
various IRQFP regimes let us study
first that of the auxiliary functions given in \eq{et} -- \eq{ekappa}.  
To start with, we take for illustration the case of $e_t$. 
Denoting by $e_i^{\infty}$'s the limit of the $e_i$'s when $Y^0_{t,b,\tau}$
tend to infinity, 
and assuming that $t_0$ is sufficiently large so
that $1+6Y^0_b\int^{t_0}_0 u_b$ is well approximated 
by $6Y^0_b\int^{t_0}_0 u_b^{\infty}$, Equation \eq{et} reads
%-\eq{ekappa} can be written 

%\beq
%e_t^{\infty}(t_0)=\frac{1}{E_t}\frac{d\tilde E_t}{d \eta}(t_0)
%+Y_b^0\frac{A_b^0 \int^{t_0}_0 u_b^{\infty}
% -\int^{t_0}_0 u_b^{\infty}e_b^{\infty}}
%{1+6Y^0_b\int^{t_0}_0 u_b^{\infty}}
%+Y_{\lambda}^0\frac{A_{\lambda}^0 \int^{t_0}_0 u_{\lambda}^{\infty}
% -\int^{t_0}_0 u_{\lambda}^{\infty}e_{\lambda}^{\infty}}
%4{1+4Y^0_{\lambda}\int^{t_0}_0 u_{\lambda}^{\infty}}
%\eeq

\beq
e_t^{\infty}(t_0)=\frac{1}{E_t}\frac{d\tilde E_t}{d \eta}(t_0)
+\frac{1}{6} A_b^0 -\frac{\int^{t_0}_0 u_b^{\infty}e_b^{\infty}}
{6 \int^{t_0}_0 u_b^{\infty}}
+Y_{\lambda}^0\frac{A_{\lambda}^0 \int^{t_0}_0 u_{\lambda}^{\infty}
 -\int^{t_0}_0 u_{\lambda}^{\infty}e_{\lambda}^{\infty}}
{1+4Y^0_{\lambda}\int^{t_0}_0 u_{\lambda}^{\infty}}
\eeq
%where $\alpha^b_t=\frac{1}{6}$ is a constant coefficient, completely
%independant of $t_0$. 

The above equation is valid for any
of the four regimes considered in section 3.2.
 To write more specific
forms for each regime one uses Eq.\eq{uFP} together with
the corresponding values of the powers given in 
the table of Appendix C. One then obtains

\beq
e^{\infty}_t=\frac{1}{E_t}\frac{d\tilde E_t}{d\eta}
-\frac{\int u^{\infty}_be^{\infty}_b}{6 \int u^{\infty}_b}
+\frac{A^0_b}{6}
+
\left\{
\begin{array}{ll}
0&
\mathrm{regimes~}1,\mathrm{~} 2 \mathrm{~and~} 4  
\cr \cr
-\frac{\int u^{\infty}_{\lambda}e^{\infty}_{\lambda}}
{4 \int u^{\infty}_{\lambda}}
+\frac{A^0_{\lambda}}{4} 
&
\mathrm{regime~}3
\end{array}
\right .
\label{eq:einftyt}
\eeq

In deriving the above equation for regimes 1, 2 and 4 we have made the implicit
assumption that the magnitude of $e_{\lambda}^{\infty}$ remains under control 
so that $\int u^{\infty}_{\lambda}e^{\infty}_{\lambda} \to 0$ with growing
$Y^0$.   

%We can extend this the same way, to all the $e_i$.

Equations \eq{eb} -- \eq{ekappa} yield along the same lines
the corresponding $e^\infty$ in the various IRQFP regimes: 
 
\beq
e^{\infty}_b=\frac{1}{E_b}\frac{d\tilde E_b}{d\eta}
-\frac{\int u^{\infty}_te^{\infty}_t}{6 \int u^{\infty}_t}
-\frac{\int u^{\infty}_{\tau}e^{\infty}_{\tau}}
{4 \int u^{\infty}_{\tau}}
+\frac{A^0_t}{6}+\frac{A^0_{\tau}}{4}
+
\left\{
\begin{array}{ll}
0
&
\mathrm{regimes~}1,\mathrm{~} 2 \mathrm{~and~} 4  
\cr \cr
-\frac{\int u^{\infty}_{\lambda}e^{\infty}_{\lambda}}
{4 \int u^{\infty}_{\lambda}}
+\frac{A^0_{\lambda}}{4}
&
\mathrm{regime~}3
\end{array}
\right .
\label{eq:einftyb}
\eeq

\beq
e^{\infty}_{\tau}=\frac{1}{E_t}\frac{d\tilde E_t}{d\eta}
-\frac{\int u^{\infty}_be^{\infty}_b}{2 \int u^{\infty}_b}
+\frac{A^0_b}{2}
+
\left\{
\begin{array}{ll}
0
&
\mathrm{regimes~}1,\mathrm{~} 2 \mathrm{~and~} 4  
\cr \cr
-\frac{\int u^{\infty}_{\lambda}e^{\infty}_{\lambda}}
{4 \int u^{\infty}_{\lambda}}
+\frac{A^0_{\lambda}}{4}
&
\mathrm{regime~}3
\end{array}
\right .
\eeq

\bea
&
e^{\infty}_{\lambda}=
\frac{1}{E_{\lambda}}\frac{d\tilde E_{\lambda}}{d\eta}
-\frac{\int u^{\infty}_te^{\infty}_t}{2 \int u^{\infty}_t}
-\frac{\int u^{\infty}_be^{\infty}_b}{2 \int u^{\infty}_b}
-\frac{\int u^{\infty}_{\tau}e^{\infty}_{\tau}}
{4\int u^{\infty}_{\tau}}
%-\frac{2Y^0_{\kappa}\int e^{\infty}_{\kappa}}
%{1+6Y^0_{\kappa}t}
&
\nonumber
\\
&
+\frac{A^0_t}{2}+\frac{A^0_b}{2}+\frac{A^0_{\tau}}{4}
+
\left\{
\begin{array}{ll}
 \frac{ 2Y^0_{\kappa} A^0_{\kappa}t  
 -2Y^0_{\kappa} \int e^{\infty}_{\kappa}}
{1+6Y^0_{\kappa}t}
&
\mathrm{regimes~}1,\mathrm{~} 2 \mathrm{~and~} 4  
\cr \cr
0
&
\mathrm{regime~}3
\end{array}
\right .
& \label{eq:einftylambda}
\eea

\beq
e^{\infty}_{\kappa}=\frac{1}{E_{\kappa}}\frac{d\tilde E_{\kappa}}{d\eta} 
+
\left\{
\begin{array}{ll}
0
&
\mathrm{regimes~}1,\mathrm{~} 2 \mathrm{~and~} 4  
\cr \cr
-\frac{3\int u^{\infty}_{\lambda}e^{\infty}_{\lambda}}
{2 \int u^{\infty}_{\lambda}}
+\frac{3A^0_{\lambda}}{2}
&
\mathrm{regime~}3
\end{array}
\right .
\label{eq:einftykappa}
\eeq

\noindent
where we used Eq.\eq{ukappaFP} when convenient. 
At this level we should stress that the particular structure of 
the above equations  will actually allow 
to solve {\sl exactly} for the dependence of the $e^\infty$'s on the initial
conditions $A_i^0$. Indeed, on one hand the $A_i^0$'s enter linearly
the inhomogeneous parts of the integral system of equations 
\eq{einftyt} -- \eq{einftykappa} with $t$ independent coefficients.
On the other hand, all the integrated parts which induce iteratively a
dependence on the $A$'s are of the form
$\frac{\int u^{\infty}_{i}e^{\infty}_{i}}
{\int u^{\infty}_{i}}$ so that any substitution therein
of the form $e_i^{\infty} \to \sum_{j} c_i^jA^0_j$, where
the $c_i^j$'s are scale independent constants,  will yield again
the same kind of dependence. It is crucial that all the coefficients
multiplying the $A$'s remain scale independent at any order of iteration of 
the integral system of equations. It follows that one can re-sum the ensuing
infinite series giving the numerical coefficients of the $A^0$'s.
 Equivalently, one can replace formally everywhere
$\frac{\int u^{\infty}_{i}e^{\infty}_{i}}
{\int u^{\infty}_{i}}$ by $e^{\infty}_{i}$ and solve the resulting {\sl linear}
system in the $e^{\infty}_{i}$'s to obtain the exact linear dependence on
the $A$'s.
The $e^{\infty}_{i}$'s take then the following form 
\beq
e_i^{\infty}=e_i^{\mathrm{FP}}+\sum_{j} \alpha_i^jA^0_j
\label{eq:eansatz}
\eeq
where the $e_i^{\mathrm{FP}}$'s are completely independent of the $A_j^0$'s.
The various coefficients $\alpha_i^j$ corresponding to the full resummation
have been summarized in Table 2.
\noindent
Note that there are in Eq.\eq{einftylambda}, regimes 1, 2, 4,  two terms
not respecting the convenient structure. Nonetheless, due to a conspiracy among
the various regimes, these terms do not invalidate
the procedure described above for the determination of the coefficients 
$\alpha_i^j$, even though $\alpha_\lambda^\kappa$ is scale dependent in 
regimes 1, 2, 4 (see Table 2).

\begin{table}
\small{
$$\vbox{\offinterlineskip \halign{
\tv# & \cc{#} & \tv# & \cc{#}  & \tv# &
\cc{#} & \tv# & \cc{#} & \tv# & \cc{#}& \tv# & \cc{#}& \tv# \cr
\noalign{\hrule}
& $\alpha_i^j$ &&\cc{$A^0_t$} &&\cc{$A^0_b$}
&&\cc{$A^0_\tau$} &&\cc{$A^0_\lambda$}
&&\cc{$A^0_\kappa$}&\cr
\noalign{\hrule}
&$e_t^\infty$&&\cc{$-\frac{2
}{61}\mathrm{~}(-\frac{5}{31})$} &&\cc{$\frac{12}{61}
\mathrm{~}(\frac{3}{31})$}
&&\cc{$-\frac{3}{61}
\mathrm{~}(-\frac{3}{31})$}
&&\cc{$0\mathrm{~}(\frac{9}{31})$}
&&\cc{$0\mathrm{~}(0)$}&\cr
\noalign{\hrule}
&\cc{$e_b^\infty$}&&\cc{$\frac{12}{61}
\mathrm{~}(\frac{3}{31})$} 
&&\cc{$-\frac{11}{61}
\mathrm{~}(-\frac{8}{31})$}
&&\cc{$\frac{18}{61}
\mathrm{~}(\frac{8}{31})$}
& &\cc{$0
\mathrm{~}(\frac{7}{31})$}
&&\cc{$0\mathrm{~}(0)$}&\cr
\noalign{\hrule}
&\cc{$e_{\tau}^\infty$}&&\cc{$-\frac{6}{61}
\mathrm{~}(-\frac{6}{31})$}
 &&\cc{$\frac{36}{61}
\mathrm{~}(\frac{16}{31})$}
&&\cc{$-\frac{9}{61}
\mathrm{~}(-\frac{17}{93})$}
& &\cc{$0
\mathrm{~}(\frac{20}{93})$}
&&\cc{$0\mathrm{~}(0)$}&\cr
\noalign{\hrule}
&\cc{$e_{\lambda}^\infty$}
&&\cc{$\frac{27}{61}\mathrm{~}(\frac{18}{31})$} 
&&\cc{$\frac{21}{61}\mathrm{~}(\frac{14}{31})$
}&&\cc{$\frac{10}{61}\mathrm{~}(\frac{20}{93})$}
& &\cc{$0\mathrm{~}(-\frac{29}{93})$}
&&\cc{$\frac{2Y^0_{\kappa}t}{1+6Y^0_{\kappa}t}\mathrm{~}(0)$}&\cr
\noalign{\hrule}
&\cc{$e_{\kappa}^\infty$}
&&\cc{$0\mathrm{~}(-\frac{27}{31})$} 
&&\cc{$0\mathrm{~}(-\frac{21}{31})$
}&&\cc{$0\mathrm{~}(-\frac{10}{31})$}
& &\cc{$0\mathrm{~}(\frac{61}{31})$}
&&\cc{$0\mathrm{~}(1)$}&\cr
\noalign{\hrule}
}}$$
}
\caption{ Resummed linear weights $\alpha_i^j$ of the $A^0$'s in the various 
$e^\infty_i$'s, {\sl i.e} $ \sum_{j} \alpha_i^jA^0_j \subset e^\infty_i $.
The unbracketed numbers are common to regimes 1,2,4 and the numbers
between brackets correspond to regime 3.}
\end{table} 

\noindent
As for the $e_i^{\mathrm{FP}}$'s, their defining equations are the same as 
those for the $e_i^\infty$'s Eqs.\eq{einftyt} -- \eq{einftykappa} with all
the $A^0$'s dropped out (the latter cancel automatically when Eq.\eq{eansatz}
is used). 

We have now all the ingredients to determine the dependence
of the various $A_i^{\infty}$ on their initial conditions. Since in all four
regimes $Y^0_{t, b, \tau}$ are large $A_{i=t, b, \tau}^{\infty}$ take
the form
$-e^{\infty}_i+\frac{\int u^{\infty}_i e^{\infty}_i}{\int u^{\infty}_i}=
-e^{\mathrm{FP}}_i+\frac{\int u^{\mathrm{FP}}_i e^{\mathrm{FP}}_i}
{\int u^{\mathrm{FP}}_i}$, see Eqs.\eq{S}, \eq{eansatz}, and are completely 
insensitive to the initial conditions $A_i^0$ similarly to the MSSM case 
\cite{Moultaka}. In contrast $A_\lambda, A_\kappa$ will be both sensitive to 
initial conditions in some regimes. 
Using Tables 1 and 2, equations \eq{S}, \eq{uansatz}, 
\eq{eansatz} one finds in the regimes
$\left \{
\begin{array}{ll}
{\mathrm{[1, 2, 4]}}
\cr
{\mathrm{[3]}}
\end{array}
\right .$

\beq
A^{\mathrm{\infty}}_{\lambda}=A^{\mathrm{FP}}_{\lambda}
-
\left \{
\begin{array}{ll}
\frac{27}{61}
\cr
0
\end{array}
\right .
A^0_t
-
\left \{
\begin{array}{ll}
\frac{21}{61}
\cr
0
\end{array}
\right .
A^0_b
-
\left \{
\begin{array}{ll}
\frac{10}{61}
\cr
0
\end{array}
\right .
A^0_{\tau}
+
\left \{
\begin{array}{ll}
1
\cr
0
\end{array}
\right .
A^0_{\lambda}
-
\left \{
\begin{array}{ll}
\frac{2Y^0_{\kappa}t}{1+6Y^0_{\kappa}t} A^0_{\kappa} \mathrm{~~} \mathrm{[1]}
\cr
\frac{1}{3} A^0_{\kappa} \mathrm{~~~~~} \mathrm{[2,4]}
\cr
0 \mathrm{~~~~~} \mathrm{[3]}
\end{array}
\right .
\label{eq:Ainftylambda}
\eeq
and
\beq
A^{\infty}_{\kappa}=A^{\mathrm{FP}}_{\kappa}
+
\left \{
\begin{array}{ll}
0
\cr
\frac{27}{31}
\end{array}
\right .
A^0_t+
\left \{
\begin{array}{ll}
0
\cr
\frac{21}{31}
\end{array}
\right .
A^0_b+
\left \{
\begin{array}{ll}
0
\cr
\frac{10}{31}
\end{array}
\right .
A^0_{\tau}
-
\left \{
\begin{array}{ll}
0
\cr
\frac{61}{31}
\end{array}
\right .
A^0_{\lambda}
+
\left \{
\begin{array}{ll}
\frac{1}{1+6Y^0_{\kappa}t} A^0_{\kappa} \mathrm{~~} \mathrm{[1]}
\cr
0 \mathrm{~~~~~} \mathrm{[2,4]}
\cr
A^0_{\kappa} \mathrm{~~~~~} \mathrm{[3]}
\end{array}
\right .
\label{eq:Ainftykappa}
\eeq

\noindent
where $A_i^{\mathrm{FP}}$ is defined in Eq.\eq{defAFP}

Hence, when one goes beyond the MSSM it is important to distinguish
between $e_i^{\mathrm{FP}}$ and $e^{\infty}_i$, the former being
useful intermediate functions to define the initial condition blind
parts of the $A_k$ in some IRQFP regimes.

\noindent
Finally let us note that the sensitivity to the initial conditions in 
Eqs.\eq{Ainftykappa}, \eq{Ainftylambda} does not imply that the physics
is no more blind to these conditions. As was already stressed in 
section 3.2, the Yukawa couplings that multiply $A_{\lambda,\kappa}$ in the
Lagrangian are vanishing in the corresponding IRQFP regimes so that
the expected screening properties are always recovered at the level of the 
physical quantities.

\vskip.5truecm
\subsection *{The $\Sigma$'s}
 
The study of the auxiliary functions $\xi_i$ is technically
more complicated than that of the $e_i$'s. The discussion goes
along the same lines as in the previous section in that
the scale independent initial conditions contributions can be easily resummed,
but there also appear scale dependent contributions in the $\xi_i$'s which we 
should discuss carefully. To illustrate the case let us consider for
simplicity the top/bottom sector switching off momentarily all other
contributions. In this case Eq.\eq{xit} reads in the limit 
$Y^0_{t,b} \to \infty$

%Now along the same Quasi-Fixed-point-regimes some simplifications
%upon the $\Sigma$'s appear. The results are reproduced for $\Sigma_t$,
%bearing in mind that it works equally for $\Sigma_b,\Sigma_\tau$ and
%in  some specific  regimes for $\Sigma_\lambda$ and $\Sigma_\kappa$. 
%If we only
%concentrate on the dependence in $\xi_b$ and $\xi_\tau$ in the 
%recursive equation leading to $\xi_t$, it gives, in the limit
%$Y^0_{t,b,\tau} \to \infty$

\bea 
 \xi_t^\infty&=&\frac{1}{E_t}\frac{d^2\tilde E_t}{d\eta d\overline{\eta}}
+\frac{2}{E_t}\frac{d\tilde E_t}{d\eta}
\left[
\frac{A_b^0}{6} -\frac{\int u_b^{\mathrm{FP}}e_b^\infty}{6\int u_b^{\mathrm{FP}}}
\right] +7\left[\frac{A_b^0}{6} -
\frac{\int u_b^{\mathrm{FP}}e_b^\infty}{6\int u_b^{\mathrm{FP}}}\right]^2
\nonumber
\\
&+&\frac{ 2A^0_b \int u_b^{\mathrm{FP}}e_b^\infty -
\int u_b^{\mathrm{FP}} \xi_b^\infty}
{6\int u_b^{\mathrm{FP}}} -\frac{1}{6} (\Sigma^0_b+(A^0_b)^2)
\label{eq:xitest}
\eea

\noindent
and similarly for Eq.\eq{xib},

\begin{equation}
\xi_b^\infty = \xi_t^\infty[ t \leftrightarrow b] \label{eq:xibest}
\end{equation} 

 Apart from the
constant terms $(\Sigma^0_{b,t} +(A^0_{b, t} )^2)$ which can be fully resummed
there is a non trivial dependence on the initial conditions in the above
equations, either explicitly or implicitly through $e_{b, t}$,
which would be difficult to handle in the iterated system \eq{xitest},
\eq{xibest}. Fortunately the corresponding terms will actually all
cancel out. To see this, it is convenient to make first the following
change of variables in Eqs.\eq{xitest}, \eq{xibest}:

\begin{equation}
\xi_k^\infty = \tilde{\xi_k}^\infty + \rho_1^k\Sigma^0_k + \rho_2^k(A^0_k)^2 +
2 \rho_3^k A^0_k e_k^\infty \label{eq:chvar}
\end{equation}

\noindent $(k = t, b)$, where $\rho_1^k, \rho_2^k$ are arbitrary constants
which can be taken equal to $1$. In the ensuing equations for 
$\tilde{\xi_t}^\infty, \tilde{\xi_b}^\infty$  we use Eq.\eq{eansatz} 
to extract all the dependence on initial conditions in $e_{b,t}^\infty$.
In the resulting equations we substitute for $\tilde{\xi_b}^\infty$ in 
$\tilde{\xi_t}^\infty$  and vice versa.
If now $\rho_3^k$ is chosen in Eq.\eq{chvar} such that
$\rho_3^{t,b} A^0_{t,b} = \sum_{j} \alpha_{t,b}^jA^0_j$, see Eq.\eq{eansatz},
and using Eqs.\eq{einftyt}, \eq{einftyb} for the reduced top/bottom system,  
one finds that all terms that contain initial conditions and
are scale dependent cancel out from the defining equations
for $\tilde{\xi_t}^\infty$ and $\tilde{\xi_b}^\infty$ 
\footnote{The specific values of
the $\alpha_{t,b}^j$ ($j=t,b$) are of course crucial for this cancellation.
It should be clear however that in the simplified example of the top/bottom 
sector we consider, these values differ from those in table 2.}.   

\noindent
Thus we are lead to a situation to the one for $e_t, e_b$ of the 
previous subsection. One can then similarly define

\bea
\tilde{\xi_k}^\infty = \xi_k^{{\mathrm FP}} +\sum_{j} \gamma_{k}^j \Sigma^0_j
+ \sum_{i,j} {\gamma_{k}^i}^j A^0_i A^0_j \label{eq:xiansatz}
\eea

\noindent
and determine the $\gamma$'s by solving a linear system of equations.
All the ingredients described here generalize to the complete
$t, b, \tau, \lambda, \kappa$ system albeit tedious algebraic
manipulations.
We will not write in this appendix the rather complicated dependence of the $\xi^{\infty}_i$ 
on the GUT scale values of the soft terms. We give here directly 
the dependence in the $\Sigma_i$'s. A useful remark is in order to understand
this dependence.  A direct calculation shows that the expression of 
$\Sigma_{i=t, b}^\infty$, Eq.\eq{S} in the limit $Y^0_i \to \infty$, is 
invariant under the substitutions
$e_i^\infty \longrightarrow\tilde e_i = e_i^\infty- \rho_3^i A^0_i, 
\xi_i^\infty \longrightarrow \tilde \xi_i,  
\tilde \xi_i \longrightarrow \xi_i^{\mathrm{FP}} $ where $\tilde \xi_i, 
\xi_i^{\mathrm{FP}}$ are as defined in Eqs.\eq{chvar}, \eq{xiansatz}. 
Equating further
the free parameter $\rho_3^i$ to $ \sum_{j} \alpha_{i}^jA^0_j$ so that
$\xi_i^{\mathrm{FP}}$ is guaranteed to be $A^0, \Sigma^0$ independent and 
$\tilde{e_i} \equiv e_i^{\mathrm{FP}}$, all dependence on the initial
conditions of the soft parameters drop out from $\Sigma_i^\infty$.
Defining

\begin{equation}
\Sigma_i^{\mathrm{FP}} = (A_i^{\mathrm{FP}})^2+
2e_i^{\mathrm{FP}}A_i^{\mathrm{FP}}
+ \xi_i^{\mathrm{FP}}- \frac{\int u_i^{\mathrm{FP}}\xi_i^{\mathrm{FP}}}{
\int u_i^{\mathrm{FP}}}
\end{equation}
one has

\beq
\Sigma_{i=t,b,\tau}^{\infty} \equiv \Sigma_i^{\mathrm{FP}}
\label{eq:sigmat}
\eeq

\beq
\Sigma_{\lambda}^{\infty}=\Sigma_{\lambda}^{\mathrm{FP}}
+\beta^j_{\lambda}\Sigma^0_j
\eeq

\noindent
and

\beq
\Sigma_{\kappa}^{\infty}=\Sigma_{\kappa}^{\mathrm{FP}}
+\beta^j_{\kappa}\Sigma^0_j
\eeq

\noindent
where the $\Sigma_{t,b,\tau}^{\mathrm{FP}}$'s are initial conditions independent
for all four IRQFP regimes
while $\Sigma_{\lambda, \kappa}^{\infty}$'s can obviously still have some 
dependence in the regimes where $Y^0_\lambda$ or $Y^0_\kappa$ are not large.
One finds for the latter in the regimes
$\left \{
\begin{array}{ll}
{\mathrm{[1, 2, 4]}}
\cr
{\mathrm{[3]}}
\end{array}
\right .$,

\bea
\Sigma^{\mathrm{\infty}}_{\lambda}&=&\Sigma^{\mathrm{FP}}_{\lambda}
-
\left \{
\begin{array}{ll}
\frac{27}{61}
\cr
0
\end{array}
\right .
\Sigma^0_t
-
\left \{
\begin{array}{ll}
\frac{21}{61}
\cr
0
\end{array}
\right .
\Sigma^0_b
-
\left \{
\begin{array}{ll}
\frac{10}{61}
\cr
0
\end{array}
\right .
\Sigma^0_{\tau}
+
\left \{
\begin{array}{ll}
1
\cr
0
\end{array}
\right .
\Sigma^0_{\lambda}
-
\left \{
\begin{array}{ll}
d^0_{\lambda} \mathrm{~~} \mathrm{[1]}
\cr
\frac{1}{3} \Sigma^0_{\kappa} \mathrm{~~~} \mathrm{[2,4]}
\cr
0 \mathrm{~~~} \mathrm{[3]}
\end{array}
\right . \label{eq:siginftylambda}
\eea
and
\beq
\Sigma^{\infty}_{\kappa}=\Sigma^{\mathrm{FP}}_{\kappa}
+
\left \{
\begin{array}{ll}
0
\cr
\frac{27}{31}
\end{array}
\right .
\Sigma^0_t+
\left \{
\begin{array}{ll}
0
\cr
\frac{21}{31}
\end{array}
\right .
\Sigma^0_b+
\left \{
\begin{array}{ll}
0
\cr
\frac{10}{31}
\end{array}
\right .
\Sigma^0_{\tau}
-
\left \{
\begin{array}{ll}
0
\cr
\frac{61}{31}
\end{array}
\right .
\Sigma^0_{\lambda}
+
\left \{
\begin{array}{ll}
d^0_{\kappa} \mathrm{~~} \mathrm{[1]}
\cr
0 \mathrm{~~~} \mathrm{[2,4]}
\cr
\Sigma^0_{\kappa} \mathrm{~~~} \mathrm{[3]}
\end{array}
\right . \label{eq:siginftykappa}
\eeq

\noindent
with 

\bea
d^0_{\lambda}=\frac{2Y^0_{\kappa}t}{1+6Y^0_{\kappa}t}
\left[\Sigma^0_{\kappa}
-\frac{(A^0_{\kappa})^2}{1+6Y^0_{\kappa}t}
\right]
\eea

\noindent 
and

\beq
d^0_{\kappa}=\frac{(A^0_{\kappa})^2}{(1+6Y^0_{\kappa}t)^2}
+\frac{\Sigma^0_{\kappa}-(A^0_{\kappa})^2}
{1+6Y^0_{\kappa}t}
\eeq

%\noindent
%for the ``corrections'' $d^0_i$, note that we can see the regime 3 
%as a limit of the regime 1 putting
%$Y^0_{\kappa}=0$, and the regime 2, as a limit of the regime 1, with
%$Y^0_{\kappa}\rightarrow \infty$.

\noindent
The $\xi^{\mathrm{FP}}$ read

\bea
&
\xt=\frac{1}{E_{t}}\frac{d^2\tilde E_{t}}{d\eta d\overline{\eta}}
-\frac{1}{E_t}\frac{d\tilde E_t}{d\eta}
\frac{\int u_b^{\fp}e_b^{\fp}}{3\int u_b^\fp}
-\frac{\int u_b^\fp \xi_b^\fp}{6 \int \ub}
+\frac{7}{36}
\left(
\frac{\int \ub \eb}{\int \ub}
\right)^2
+d^{\lambda}_t
&
\eea

\bea
&
\xb=\frac{1}{E_{b}}\frac{d^2\tilde E_{b}}{d\eta d\overline{\eta}}
-\frac{1}{E_b}\frac{d\tilde E_b}{d\eta}
\left(
\frac{\int \ut \et}{3\int \ut}
+\frac{\int \utau \etau}{2\int \utau}
\right)
-\frac{\int u_t^\fp \xi_t^\fp}{6 \int \ut}
-\frac{\int \utau \xtau}{4 \int \utau}
\nonumber
&
\\
&
+\frac{7}{36}
\left(
\frac{\int \ut \et}{\int \ut}
\right)^2
+\frac{5}{16}
\left(
\frac{\int \utau \etau}{\int \utau}
\right)^2
+\frac{1}{12}
\frac{\int \ut \et}{\int \ut}\frac{\int \utau \etau}{\int \utau}
+d^{\lambda}_b
&
\eea

\bea
&
\xtau=\frac{1}{E_{\tau}}\frac{d^2\tilde E_{\tau}}{d\eta d\overline{\eta}}
-\frac{1}{E_\tau}\frac{d\tilde E_\tau}{d\eta}
\frac{\int u_b^{\fp}e_b^{\fp}}{\int u_b^\fp}
-\frac{\int u_b^\fp \xi_b^\fp}{2 \int \ub}
+\frac{3}{4}
\left(
\frac{\int \ub \eb}{\int \ub}
\right)^2
+d^{\lambda}_\tau
&
\eea

\bea
&\xi^{\mathrm{FP}}_{\lambda}=
\frac{1}{E_{\lambda}}\frac{d^2\tilde E_{\lambda}}{d\eta d\overline{\eta}}
-\frac{1}{E_\lambda}\frac{d\tilde E_\lambda}{d\eta}
\left(
\frac{\int u_t^{\mathrm{FP}} e_t^{\mathrm{FP}}}
{\int u_t^{\mathrm{FP}}}
+\frac{\int u_b^{\mathrm{FP}} e_b^{\mathrm{FP}}}
{\int u_b^{\mathrm{FP}}}
+\frac{\int u_\tau^{\mathrm{FP}} e_\tau^{\mathrm{FP}}}
{2\int u_\tau^{\mathrm{FP}}}
\right)
\nonumber
&
\\
&
-\frac{\int u_t^{\mathrm{FP}} \xi_t^{\mathrm{FP}}}
{2\int u_t^{\mathrm{FP}}}
-\frac{\int u_b^{\mathrm{FP}} \xi_b^{\mathrm{FP}}}
{2\int u_b^{\mathrm{FP}}}
-\frac{\int u_\tau^{\mathrm{FP}} \xi_\tau^{\mathrm{FP}}}
{4\int u_\tau^{\mathrm{FP}}}
\nonumber
&
\\
&
+\frac{3}{4}
\left(
\frac{\int u_t^{\mathrm{FP}} e_t^{\mathrm{FP}}}
{\int u_t^{\mathrm{FP}}}
\right)^2
+\frac{3}{4}
\left(
\frac{\int u_b^{\mathrm{FP}} e_b^{\mathrm{FP}}}
{\int u_b^{\mathrm{FP}}}
\right)^2
+\frac{5}{16}
\left(
\frac{\int u_\tau^{\mathrm{FP}} e_\tau^{\mathrm{FP}}}
{\int u_\tau^{\mathrm{FP}}}
\right)^2
\nonumber
&
\\
&
+\frac{1}{2}
\frac{\int \ut \et}{\int \ut}\frac{\int \ub \eb}{\int \ub}
+\frac{1}{4}
\frac{\int \ut \et}{\int \ut}\frac{\int \utau \etau}{\int \utau}
+\frac{1}{4}
\frac{\int \ub \eb}{\int \ub}\frac{\int \utau \etau}{\int \utau}
+d^{\kappa}_{\lambda}
&
\eea

\bea
&
\xk=\frac{1}{E_{\kappa}}\frac{d^2\tilde E_{\kappa}}{d\eta d\overline{\eta}}
+d^{\lambda}_{\kappa}
&
\eea

\noindent
It turns out that in regimes 1, 2, and 4 one has  
$d_i^{\lambda}=0$\\

\noindent
while in regime 3, we have

\beq
d^{\lambda}_t=-\frac{2}{E_t}\frac{d\tilde E_t}{d\eta}
\frac{\int u_{\lambda}^{\mathrm{FP}} e_{\lambda}^{\mathrm{FP}}}
{4\int u_{\lambda}^{\mathrm{FP}}}
-\frac{\int u_{\lambda}^{\mathrm{FP}}\xi_{\lambda}^{\mathrm{FP}}}
{4\int u_{\lambda}^{\mathrm{FP}}}
+\frac{5}{16}
\left(
\frac{\int u_{\lambda}^{\mathrm{FP}} e_{\lambda}^{\mathrm{FP}}}
{\int u_{\lambda}^{\mathrm{FP}}}
\right)^2
+\frac{1}{12} \frac{\int u_b^{\mathrm{FP}}e_b^{\mathrm{FP}}}
{\int u_b^{\mathrm{FP}}}
\frac{\int u_{\lambda}^{\mathrm{FP}} e_{\lambda}^{\mathrm{FP}}}
{\int u_{\lambda}^{\mathrm{FP}}}
\eeq 

\beq
d^{\lambda}_b=-\frac{2}{E_b}\frac{d\tilde E_b}{d\eta}
\frac{\int u_{\lambda}^{\mathrm{FP}} e_{\lambda}^{\mathrm{FP}}}
{4\int u_{\lambda}^{\mathrm{FP}}}
-\frac{\int u_{\lambda}^{\mathrm{FP}}\xi_{\lambda}^{\mathrm{FP}}}
{4\int u_{\lambda}^{\mathrm{FP}}}
+\frac{5}{16}
\left(
\frac{\int u_{\lambda}^{\mathrm{FP}} e_{\lambda}^{\mathrm{FP}}}
{\int u_{\lambda}^{\mathrm{FP}}}
\right)^2
+\frac{\int u_{\lambda}^{\mathrm{FP}} e_{\lambda}^{\mathrm{FP}}}
{\int u_{\lambda}^{\mathrm{FP}}}
\left(
\frac{1}{12} 
\frac{\int u_t^{\mathrm{FP}}e_t^{\mathrm{FP}}}
{\int u_t^{\mathrm{FP}}}
+\frac{1}{8}
\frac{\int u_{\tau}^{\mathrm{FP}} e_{\tau}^{\mathrm{FP}}}
{\int u_{\tau}^{\mathrm{FP}}}
\right)
\eeq

\beq
d^{\lambda}_\tau=-\frac{2}{E_\tau}\frac{d\tilde E_\tau}{d\eta}
\frac{\int u_{\lambda}^{\mathrm{FP}} e_{\lambda}^{\mathrm{FP}}}
{4\int u_{\lambda}^{\mathrm{FP}}}
-\frac{\int u_{\lambda}^{\mathrm{FP}}\xi_{\lambda}^{\mathrm{FP}}}
{4\int u_{\lambda}^{\mathrm{FP}}}
+\frac{5}{16}
\left(
\frac{\int u_{\lambda}^{\mathrm{FP}} e_{\lambda}^{\mathrm{FP}}}
{\int u_{\lambda}^{\mathrm{FP}}}
\right)^2
+\frac{1}{4} \frac{\int u_b^{\mathrm{FP}}e_b^{\mathrm{FP}}}
{\int u_b^{\mathrm{FP}}}
\frac{\int u_{\lambda}^{\mathrm{FP}} e_{\lambda}^{\mathrm{FP}}}
{\int u_{\lambda}^{\mathrm{FP}}}
\eeq

For $d_\lambda^\kappa$ we find
\begin{itemize}
\item regime 1
\bea
&
d^{\kappa}_{\lambda}=
-\frac{4}{E_\kappa}\frac{d\tilde E_\kappa}{d\eta}
Y^0_{\kappa}\frac{\int  \ek}{1+6Y^0_{\kappa} t}
-2Y^0_{\kappa}\frac{\int  \xk}{1+6Y^0_{\kappa} t}
+16(Y^0_{\kappa})^2
\left(
\frac{\int  \ek}{1+6Y^0_{\kappa} t}
\right)^2
\nonumber
&
\\
&
+Y^0_{\kappa}\frac{\int  \ek}{1+6Y^0_{\kappa} t}
\left(
 2\frac{\int \ut \et}{\int \ut}
+2\frac{\int \ub \eb}{\int \ub}
+\frac{\int \utau \etau}{\int \utau}
\right)
\eea

\item regimes 2 and 4

\bea
&
d^{\kappa}_{\lambda}=
-\frac{2}{E_\kappa}\frac{d\tilde E_\kappa}{d\eta}
\frac{\int  \ek}{3 t}
-\frac{\int  \xk}{3 t}
+\frac{4}{9}
\left(
\frac{\int  \ek}{ t}
\right)^2
\nonumber
&
\\
&
+\frac{\int  \ek}{6 t}
\left(
 2\frac{\int \ut \et}{\int \ut}
+2\frac{\int \ub \eb}{\int \ub}
+\frac{\int \utau \etau}{\int \utau}
\right)
\eea

\item regime 3

\beq
d^{\kappa}_{\lambda}=0
\eeq

\end{itemize}

and for $d^\lambda_{\kappa}$
\beq
d^\lambda_{\kappa}=
\left\{
\begin{array}{ll}
0 
&
\mathrm{regimes~}1, 2, \mathrm{~and~} 4
\cr \cr
-\frac{3}{E_\kappa}\frac{d\tilde E_\kappa}{d\eta}
\frac{\int \ul \el}{\int \ul}
-\frac{3}{2}\frac{\el \xl}{\int \ul}
+\frac{15}{4}\frac{\int \ul \el}{\int \ul}
&
\mathrm{regime~}3
\cr
\end{array}
\right.
\eeq

%\section*{Appendix D: Soft terms dependence on the initial conditions}
%\renewcommand{\theequation}{D.\arabic{equation}}
%\setcounter{equation}{0}

\vskip.5truecm
\subsection*{The $m$'s.}

When we develop the expressions (B16-B23) in the different FP regimes
and replace the $\Sigma_t$ by their dependence on the soft masses
$m^2_i$, we find solutions of the form

\begin{equation}
m_i^2=(m_i^2)_{m^0}+(m_i^2)_{M^0}+(m_i^2)_{Tr}
\end{equation}

\noindent
where $(m_i^2)_{Tr}$ is given by \eq{mtrace}
and the $(m_i^2)_{m^0}$ are written below.
But, there is no compelling exact analytical solutions for the
$(m_i^2)_{M^0}$.
 For the regimes 1,2 and 4 :

\bea
(m_{Q})^2_{m^0}
&=&\frac{85}{122}m^{0^{\;2}}_{Q_3}-\frac{17}{122}m^{0^{\;2}}_{U_3}-\frac{10}{61}m^{0^{\;2}}_{D_3}
+\frac{5}{122}m^{0^{\;2}}_{L}+\frac{5}{122}m_{E_3}^{0^{\;2}}
-\frac{15}{122}m_1^{0^{\;2}}-\frac{17}{122}m_2^{0^{\;2}}
\nonumber
\\
\nonumber
\\
(m_{T})^2_{m^0}
&=&-\frac{17}{61}m^{0^{\;2}}_{Q_3}+\frac{40}{61}m^{0^{\;2}}_{U_3}+\frac{4}{61}m^{0^{\;2}}_{D_3}
-\frac{1}{61}m^{0^{\;2}}_{L}-\frac{1}{61}m_{E_3}^{0^{\;2}}
+\frac{3}{61}m_1^{0^{\;2}}-\frac{21}{61}m_2^{0^{\;2}}
\nonumber
\\
\nonumber
\\
(m_{B})^2_{m^0}
&=&-\frac{20}{61}m^{0^{\;2}}_{Q_3}+\frac{4}{61}m^{0^{\;2}}_{U_3}+\frac{37}{61}m^{0^{\;2}}_{D_3}
+\frac{6}{61}m^{0^{\;2}}_{L}+\frac{6}{61}m_{E_3}^{0^{\;2}}
-\frac{18}{61}m_1^{0^{\;2}}+\frac{4}{61}m_2^{0^{\;2}}
\nonumber
\\
\nonumber
\\
(m_{L})^2_{m^0}
&=&\frac{15}{122}m^{0^{\;2}}_{Q_3}-\frac{3}{122}m^{0^{\;2}}_{U_3}+\frac{9}{61}m^{0^{\;2}}_{D_3}
+\frac{87}{122}m^{0^{\;2}}_{L}-\frac{35}{122}m_{E_3}^{0^{\;2}}
-\frac{17}{122}m_1^{0^{\;2}}-\frac{3}{122}m_2^{0^{\;2}}
\nonumber
\\
\nonumber
\\
(m_{E})^2_{m^0}
&=&\frac{15}{61}m^{0^{\;2}}_{Q_3}-\frac{3}{61}m^{0^{\;2}}_{U_3}+\frac{18}{61}m^{0^{\;2}}_{D_3}
-\frac{35}{61}m^{0^{\;2}}_{L}+\frac{26}{61}m_{E_3}^{0^{\;2}}
-\frac{17}{61}m_1^{0^{\;2}}-\frac{3}{61}m_2^{0^{\;2}}
\nonumber
\\
\nonumber
\\
(m_1)^2_{m^0}
&=&-\frac{45}{122}m^{0^{\;2}}_{Q_3}+\frac{9}{122}m^{0^{\;2}}_{U_3}-\frac{27}{61}m^{0^{\;2}}_{D_3}
-\frac{17}{122}m^{0^{\;2}}_{L}-\frac{17}{122}m_{E_3}^{0^{\;2}}
+\frac{51}{122}m_1^{0^{\;2}}+\frac{9}{122}m_2^{0^{\;2}}
\nonumber
\\
\nonumber
\\
(m_2)^2_{m^0}
&=&-\frac{51}{122}m^{0^{\;2}}_{Q_3}-\frac{63}{122}m^{0^{\;2}}_{U_3}+\frac{6}{61}m^{0^{\;2}}_{D_3}
-\frac{3}{122}m^{0^{\;2}}_{L}-\frac{3}{122}m_{E_3}^{0^{\;2}}
+\frac{9}{122}m_1^{0^{\;2}}+\frac{59}{122}m_2^{0^{\;2}}
\nonumber
\\
\nonumber
\\
(m_{S})^2_{m^0}&=&
\left \{
\begin{array}{ll}
\frac{m^{0^{\;2}}_S+396Y^0_{\kappa}m^{0^{\;2}}_S-132Y^0_{\kappa}(A^0_{\kappa})^2}
{(1+396Y^0_{\kappa})^2} 
\mathrm{~~~} \mathrm{regime} \mathrm{~~~} 1
\cr
m^{0^{\;2}}_S \mathrm{~~~~~~~~~~~~~~~~~~~~~~~~~~~~~} 
\mathrm{regime} \mathrm{~~~} 2,\mathrm{~} 4
\end{array}
\right . \label{eq:msoft124}
\eea

In the regime 3. we obtained

\bea
(m_{Q})^2_{m^0}
&=&\frac{39}{62}m^{0^{\;2}}_{Q_3}-\frac{11}{62}m^{0^{\;2}}_{U_3}-\frac{6}{31}m^{0^{\;2}}_{D_3}
+\frac{5}{186}m^{0^{\;2}}_{L}+\frac{5}{186}m_{E_3}^{0^{\;2}}
-\frac{5}{62}m_1^{0^{\;2}}-\frac{17}{186}m_2^{0^{\;2}}+\frac{8}{93}m^{0^{\;2}}_S
\nonumber
\\
\nonumber
\\
(m_{T})^2_{m^0}
&=&-\frac{11}{31}m^{0^{\;2}}_{Q_3}+\frac{19}{31}m^{0^{\;2}}_{U_3}+\frac{1}{31}m^{0^{\;2}}_{D_3}
-\frac{1}{31}m^{0^{\;2}}_{L}-\frac{1}{31}m_{E_3}^{0^{\;2}}
+\frac{3}{31}m_1^{0^{\;2}}-\frac{9}{31}m_2^{0^{\;2}}+\frac{3}{31}m^{0^{\;2}}_S
\nonumber
\\
\nonumber
\\
(m_{B})^2_{m^0}
&=&-\frac{12}{31}m^{0^{\;2}}_{Q_3}+\frac{1}{31}m^{0^{\;2}}_{U_3}+\frac{18}{31}m^{0^{\;2}}_{D_3}
+\frac{8}{93}m^{0^{\;2}}_{L}+\frac{8}{93}m_{E_3}^{0^{\;2}}
-\frac{8}{31}m_1^{0^{\;2}}+\frac{10}{93}m_2^{0^{\;2}}+\frac{7}{93}m^{0^{\;2}}_S
\nonumber
\\
\nonumber
\\
(m_{L})^2_{m^0}
&=&\frac{5}{62}m^{0^{\;2}}_{Q_3}-\frac{3}{62}m^{0^{\;2}}_{U_3}+\frac{4}{31}m^{0^{\;2}}_{D_3}
+\frac{131}{186}m^{0^{\;2}}_{L}-\frac{55}{186}m_{E_3}^{0^{\;2}}
-\frac{7}{62}m_1^{0^{\;2}}+\frac{1}{186}m_2^{0^{\;2}}+\frac{5}{93}m^{0^{\;2}}_S
\nonumber
\\
\nonumber
\\
(m_{E})^2_{m^0}
&=&\frac{5}{31}m^{0^{\;2}}_{Q_3}-\frac{3}{31}m^{0^{\;2}}_{U_3}+\frac{8}{31}m^{0^{\;2}}_{D_3}
-\frac{55}{93}m^{0^{\;2}}_{L}+\frac{38}{93}m_{E_3}^{0^{\;2}}
-\frac{7}{31}m_1^{0^{\;2}}-\frac{1}{93}m_2^{0^{\;2}}+\frac{10}{93}m^{0^{\;2}}_S
\nonumber
\\
\nonumber
\\
(m_1)^2_{m^0}
&=&-\frac{15}{62}m^{0^{\;2}}_{Q_3}+\frac{9}{62}m^{0^{\;2}}_{U_3}-\frac{12}{31}m^{0^{\;2}}_{D_3}
-\frac{7}{62}m^{0^{\;2}}_{L}-\frac{7}{62}m_{E_3}^{0^{\;2}}
+\frac{21}{62}m_1^{0^{\;2}}-\frac{1}{62}m_2^{0^{\;2}}-\frac{5}{31}m^{0^{\;2}}_S
\nonumber
\\
\nonumber
\\
(m_2)^2_{m^0}
&=&-\frac{17}{62}m^{0^{\;2}}_{Q_3}-\frac{27}{62}m^{0^{\;2}}_{U_3}+\frac{5}{31}m^{0^{\;2}}_{D_3}
+\frac{1}{186}m^{0^{\;2}}_{L}+\frac{1}{186}m_{E_3}^{0^{\;2}}
-\frac{1}{62}m_1^{0^{\;2}}+\frac{71}{186}m_2^{0^{\;2}}-\frac{17}{93}m^{0^{\;2}}_S
\nonumber
\\
\nonumber
\\
(m_{S})^2_{m^0}
&=&\frac{16}{31}m^{0^{\;2}}_{Q_3}+\frac{9}{31}m^{0^{\;2}}_{U_3}+\frac{7}{31}m^{0^{\;2}}_{D_3}
+\frac{10}{93}m^{0^{\;2}}_{L}+\frac{10}{93}m_{E_3}^{0^{\;2}}
-\frac{10}{31}m_1^{0^{\;2}}-\frac{34}{93}m_2^{0^{\;2}}+\frac{32}{93}m^{0^{\;2}}_S
\nonumber
\\ \label{eq:msoft3}
\eea

\end{document}